\documentclass[11pt]{article}

\usepackage[final]{acl}

\usepackage{times}
\usepackage{latexsym}
\usepackage{marvosym}       
\usepackage{microtype}      
\usepackage{xcolor}         

\usepackage[utf8]{inputenc} 
\usepackage[T1]{fontenc}    
\usepackage{hyperref}       
\usepackage{url}            
\usepackage{booktabs}       
\usepackage{amsfonts}       
\usepackage{nicefrac}       

\usepackage{colortbl}
\usepackage{enumitem}
\usepackage{amsmath,amsfonts,bm,amssymb}
\usepackage{wrapfig}
\usepackage{makecell,multirow}

\usepackage{color}
\definecolor{purple}{RGB}{160,32,240}

\usepackage{graphicx}
\usepackage{caption}
\usepackage{subcaption}
\usepackage{dblfloatfix}
\usepackage{placeins}
\usepackage{algorithm}
\usepackage{algpseudocode}
\usepackage{float}
\usepackage{threeparttable}

\definecolor{perfup}{HTML}{2E8B57}
\definecolor{cfgcolhl}{RGB}{243,247,252}

\usepackage{color}
\newcommand{\diff}[1]{{\color{blue}#1}} 

\usepackage{subfiles}
\usepackage[normalem]{ulem}

\usepackage[most]{tcolorbox}
\newtcolorbox{insightbox}{
  enhanced,
  colback=black!2,
  colframe=black!20,
  boxrule=0.4pt,
  arc=2pt,
  left=6pt,
  right=6pt,
  top=4pt,
  bottom=4pt,
  boxsep=0pt,
  before skip=6pt,
  after skip=6pt
}

\newtcbox{\insightinlinebox}{
  on line,
  enhanced,
  colback=black!2,
  colframe=black!20,
  boxrule=0.4pt,
  arc=2pt,
  left=6pt,
  right=6pt,
  top=4pt,
  bottom=4pt,
  boxsep=0pt,
  before skip=6pt,
  after skip=6pt
  nobeforeafter,
}

\usepackage[T1]{fontenc}

\usepackage[utf8]{inputenc}

\usepackage{microtype}

\usepackage{inconsolata}

\usepackage{graphicx}
\usepackage{afterpage}

\title{Unlocking Multimodal Protein Language Models at Inference Time}

\newcommand{\correspondingauthor}{%
  \begingroup
  \renewcommand{\thefootnote}{\Letter}%
  \footnotemark
  \endgroup}

\author{
 \textbf{Yi Zhou\textsuperscript{1}},
 \textbf{Qipeng Wang\textsuperscript{1}},
 \textbf{Yunqing Liu\textsuperscript{1}},
 \textbf{Jun Xia\textsuperscript{2}},
 \textbf{Qing Li\textsuperscript{1}},
 \textbf{Wenqi Fan\textsuperscript{1,}\correspondingauthor},
\\
 \textsuperscript{1}The Hong Kong Polytechnic University, \\
 \textsuperscript{2}The Hong Kong University of Science and Technology (Guangzhou),
\\
 \small{
   \href{mailto:echo-yi.zhou@connect.polyu.hk}{echo-yi.zhou@connect.polyu.hk}; \href{mailto:qipeng26.wang@connect.polyu.hk}{qipeng26.wang@connect.polyu.hk}; 
   \href{mailto:yunqing617.liu@connect.polyu.hk}{yunqing617.liu@connect.polyu.hk};}\\
\small{
   \href{mailto:junxia@hkust-gz.edu.cn}{junxia@hkust-gz.edu.cn}; 
   \href{mailto:csqli@comp.polyu.edu.hk}{csqli@comp.polyu.edu.hk}; 
   \href{mailto:wenqifan03@gmail.com}{wenqifan03@gmail.com}
 }
}

\begin{document}
\maketitle
\begingroup
\renewcommand{\thefootnote}{\Letter}
\footnotetext{Corresponding author: Wenqi Fan.}
\endgroup
\begin{abstract}
Multimodal protein language models (pLMs) learn joint protein sequence-structure distributions, and their generation performance should also depend critically on inference-time sampling strategies.
Yet prior work has focused more on model training than on how inference-time strategies behave.
In this paper, we establish a three-stage investigation framework to empirically study the inference design space of multimodal pLMs across three representative pLMs and four fundamental tasks.
We evaluate vanilla sampling, task-specific classifier-free guidance, and reward-guided beam search on multimodal pLMs, corresponding to controls over sampling distributions, per-step logits, and parallel trajectories.
Throughout the complementary advancements centered on exploration-exploitation trade-off, we (1) reveal the suboptimality of default inference protocols and identify task-oriented sampling preferences; (2) observe substantial quantitative gains across tasks, consistently boosting the upper bound performance of multimodal pLMs without updating model parameters; (3) derive conclusions about base models that differ from prior consensus.
Our code is available at: \href{https://github.com/EchoChou990919/mplm_inference}{github.com/EchoChou990919/mplm\_inference}.

\end{abstract}

\section{Introduction}
 
Proteins are essential macromolecules in virtually all cellular processes.
Their behavior follows the sequence-structure-function paradigm~\citep{anfinsen1973principles} -- the amino acid sequence determines the 3D structure, which in turn defines biological function. 
This biological centrality, combined with the inherent challenge of aligning heterogeneous data modalities within a unified semantic space, has driven extensive research on computational protein modeling~\cite{fan2025computational,notin2024machine,liu2026geometric,liu-etal-2025-glprotein}. 
Especially in the context of protein sequence-structure duality, key tasks like protein structure prediction~\cite{jumper2021highly,lin2023esm2}, inverse folding~\cite{dauparas2022robust}, motif scaffolding~\cite{watson2023novo}, and unconditional protein sequence-structure co-generation~\cite{geffner2025laproteina} have attracted sustained research interest.

In recent years, multimodal generative protein language models (\textbf{pLMs}) have emerged as an ``all-in-one'' solution for \textbf{joint protein sequence-structure modeling}~\cite{fan2025computational,liu2026enhancing}. 
Representative models, such as ESM3~\cite{hayes2025simulating} and DPLM-2 series~\cite{wang2024dplm,hsieh2025elucidating}, treat protein sequence and structure as parallel token tracks: they learn the multimodal probability distribution of proteins during pre-training and address diverse tasks via iterative sampling from pLM logits at inference time. 
As with modern language and vision generative models~\cite{karras2024guiding,snell2024scaling}, the performance of multimodal pLMs on generative tasks depends not only on the probability landscape learned but also on the sampling strategy employed, which shapes the inference-time sampling trajectory.

\begin{figure}
  \centering
  \includegraphics[width=\linewidth]{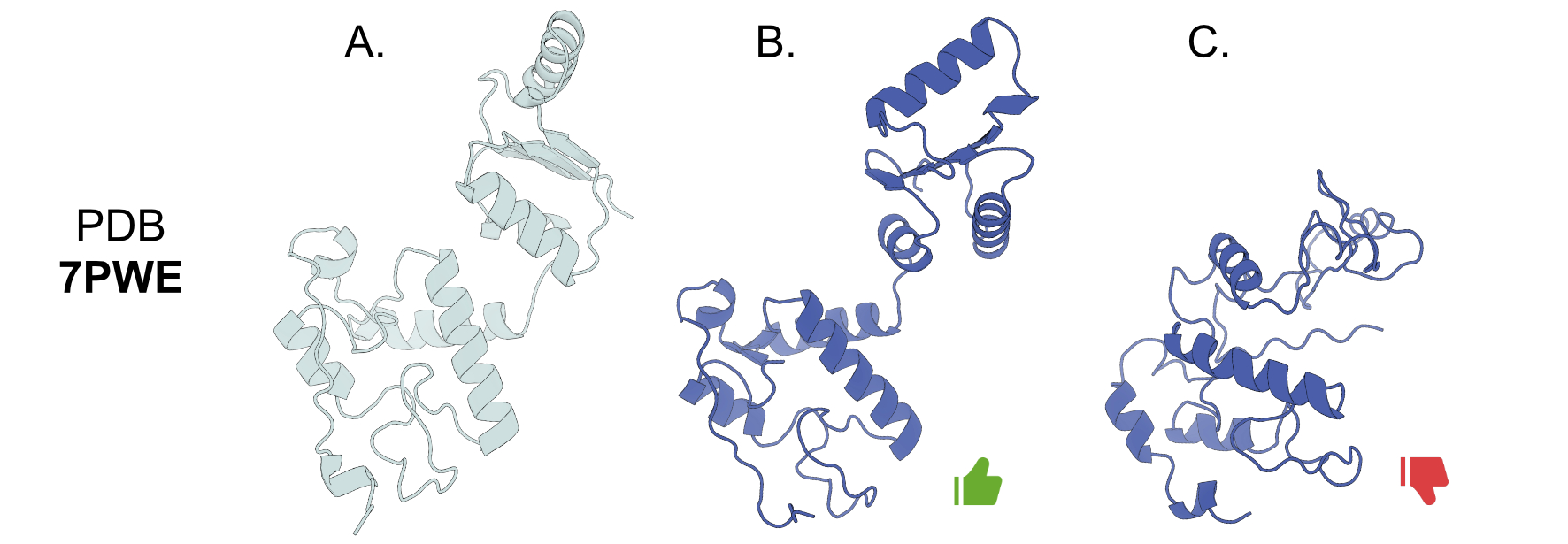}
  \captionsetup{font=small,skip=3pt}
  \caption{
  A structure prediction case. \textbf{A}. Ground-truth. \textbf{B}. One-step argmax prediction result. \textbf{C}. $L$-step temperature-based sampling result.
  }
  \vspace{-12pt}
  \label{fig:intro}
\end{figure}

Figure~\ref{fig:intro} illustrates a protein structure prediction case comparing two typical inference-time sampling configurations. 
While single-step decoding at temperature $=0.0$, ESM3 successfully predicts the structure of protein 7PWE in generally the same fold (TM-score $> 0.5$). 
However, when structure tokens are iteratively sampled at temperature $=1.0$ over $L$ steps, ESM3's final prediction obviously collapses.
This disparity intuitively underscores the significance of sampling strategies at inference time.
On the same pLM landscape and starting from the same conditional point, inference over an exploitative short trajectory or an exploratory long trajectory yields different outputs.

Nevertheless, most research on multimodal pLMs has focused primarily on the effectiveness of training and fine-tuning, leaving inference-time sampling strategies largely underexplored. 
ESM3~\cite{hayes2025simulating}, DPLM-2~\cite{wang2024dplm}, and DPLM-2.1~\cite{hsieh2025elucidating} perform protein generation tasks by default using \textit{model- and task-specific vanilla sampling configurations}, without the widely recognized \textit{reward-free guidance}~\cite{ho2022classifier,karras2024guiding,nie2024scaling} and \textit{reward-guided search}~\cite{yao2023tree,wei2025unifying} techniques. 
As a result, cross-model comparisons can be confounded by implementation-specific inference defaults.
A few recent attempts apply advanced sampling to ESM3, e.g., a self-consistency reward for inverse folding~\cite{liu2025protinvtree} or a foldability reward for unconditional generation~\cite{li2024derivative}, but each is confined to one task, a single base model, or a specific strategy.
We still lack a systematic understanding of \textit{how multi-level sampling strategies behave across multimodal pLMs and tasks}, and \textit{how the underlying exploration-exploitation trade-off should be navigated at inference time}.

\begin{figure*}[t]
    \centering
    \includegraphics[width=0.95\linewidth]{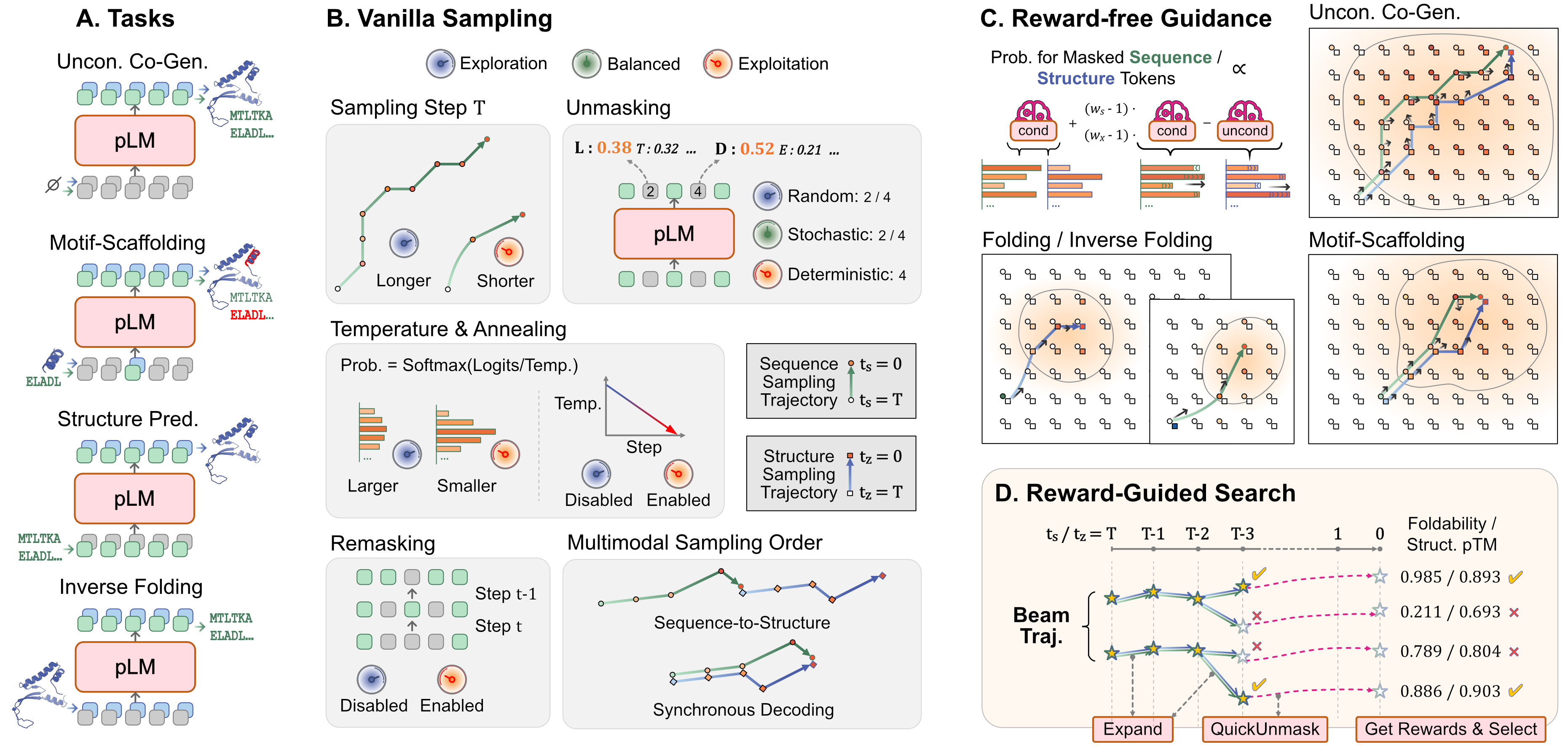}
    \captionsetup{font=small,skip=6pt}
    \caption{\textbf{Overview of our empirical investigation framework.}
        \textbf{A.} Multimodal protein language models (pLMs) perform four foundational protein modeling tasks via iterative diffusion sampling.
        \textbf{B.} At the \uline{distribution} level, we benchmark \textbf{vanilla sampling} across three pLMs and four tasks, therefore identifying the task-oriented exploration-exploitation trade-off at inference time.
        \textbf{C.} At the \uline{logit} level, we propose \textbf{task-motivated classifier-free guidance} that steers each step's logits and therefore directs the sampling trajectory.
        \textbf{D.} At the \uline{trajectory} level, \textbf{reward-guided beam search} navigates parallel trajectories using model-internal quality signals, highlighting the effectiveness of ``explore-then-select'' inference across tasks. 
    }
    \label{fig:overview}
    \vspace{-12pt}
\end{figure*}

In this paper, we address this gap through a systematic investigation that organizes \textbf{inference-time sampling} along three orthogonal control axes at different granularity levels: 
the sampling \uline{distribution}, the per-step \uline{logits}, and the parallel sampling \uline{trajectories}, with an overview presented in Figure~\ref{fig:overview}. 
\textbf{Section 3}: At the distribution level, we benchmark \textbf{vanilla sampling} among ESM3, DPLM-2, and DPLM-2.1 on four protein modeling tasks, sweeping the multimodal sampling order, total denoising step, unmasking, remasking, and temperature settings.
\textbf{Section 4}: At the logit level, we instantiate task-specific variants of \textbf{classifier-free guidance (CFG)} that redirect each step's distribution toward the condition-aligned region of the sampling space.
\textbf{Section 5}: At the trajectory level, we introduce a reward-guided \textbf{beam search} that selects across parallel sampling trajectories using global quality signals. 
Our main contributions are summarized as follows, while the main insights are summarized at the end of each section:

\begin{itemize} [leftmargin=*,topsep=0pt,itemsep=3pt,parsep=0pt]
    \item We introduce a systematic investigation framework for inference-time sampling in multimodal pLMs, orthogonally composing vanilla sampling, reward-free guidance, and reward-guided search strategies, and comprehensively covering four generative protein modeling tasks.
    \item The three-level inference strategies we investigate consistently advance the best empirical performance for multimodal pLMs across four tasks, especially boosting breakthroughs in ESM3's unconditional protein sequence-structure cogeneration and motif-scaffolding. 
    \item We demonstrate that standardized inference-time strategies can substantially change conclusions about multimodal pLMs: models viewed as weak under default sampling can become competitive once inference strategies are controlled, revealing inference protocol as a hidden confounder in protein modeling evaluation. 
\end{itemize}

\section{Preliminaries}

\noindent \textbf{Multimodal Protein Language Modeling: Formulation and Benchmarks.}
A protein is fundamentally characterized by its sequence and structure. 
We study multimodal pLMs, represented by ESM3~\cite{hayes2025simulating} and DPLM-2 series~\cite{wang2024dplm,hsieh2025elucidating}, which model proteins via parallel token tracks.
For a protein of length $L$, its sequence is represented as $\bm{s} = (s_1, \dots, s_L)$ over the standard 20-amino-acid vocabulary, and its structure as a discrete latent sequence $\bm{z} = (z_1, \dots, z_L)$ produced by the model's structure tokenizer. 
These pLMs learn the joint distribution $p_\theta(\bm{s}, \bm{z})$ with neural network $\theta$, from which four multimodal protein generative tasks can be conducted: 
unconditional sequence-structure co-generation via $p_\theta(\bm{s}, \bm{z} | \varnothing)$, motif scaffolding via $p_\theta(\bm{s}, \bm{z} | \bm{s}_{\text{motif}}, \bm{z}_{\text{motif}})$, 
structure prediction via $p_\theta(\bm{z} | \bm{s})$, and inverse folding via $p_\theta(\bm{s} | \bm{z})$.

Our study focuses on investigating inference protocols for multimodal pLMs. 
Regarding the evaluation datasets and metrics, we follow established benchmarks. 
For unconditional co-generation, we generate protein samples ranging from 100 to 500 residues and report designability and diversity metrics~\cite{geffner2025laproteina}.
Motif scaffolding is evaluated on 24 benchmark problems~\citep{zheng2025motifbench}, with the number of solved problems, overall success rate, and diversity of successful designs reported~\cite{wang2024dplm,yim2024improved}.
Protein structure prediction and inverse folding are evaluated on the CAMEO2022 and PDB date datasets~\cite{campbell2024generative}, with RMSD \& TM-score, and scTM \& AAR reported~\cite{wang2024dplm,hsieh2025elucidating}.
Additionally, we record the FLOPs and wall-clock runtime to analyze the computational efficiency of inference protocols.
Details of the protein structure tokenization, base models, and evaluation pipelines are deferred to Appendix~\ref{asec:mplm_preliminary}.

\noindent \textbf{Protein Generation with Diffusion Language Models.}
Typical multimodal pLMs~\cite{hayes2025simulating,wang2024dplm,hsieh2025elucidating} are built on the discrete diffusion framework. 
At inference, protein generation proceeds along a reverse denoising trajectory of \textbf{$T$ steps}, iteratively refining an initial masked state $(\bm{s}^{(T)}, \bm{z}^{(T)})$ into a clean state $(\bm{s}^{(0)}, \bm{z}^{(0)})$. 
With two token tracks, multimodal protein generation admits different \textbf{sampling orders} along the two diffusion timelines $t_s$ and $t_z$: a \textit{sequence-to-structure} schedule completes $t_s$ before advancing $t_z$, 
while a \textit{synchronous} schedule couples them under a shared timeline ($t_s = t_z = t$).

Unlike left-to-right autoregressive decoding, each diffusion step involves two sets of operations on the masked positions. 
First, an explicit \textbf{unmasking} policy selects an unmasking subset $\mathcal{U}_t \subseteq \mathcal{M}_t$ from the currently masked indices $\mathcal{M}_t$, based on a per-position confidence $c_i^{(t)}$ (e.g., log-probability or entropy),
in one of three modes: \textit{deterministic} (best $c_i^{(t)}$ first), \textit{stochastic} (Gumbel-perturbed confidence), or \textit{random}.
An optional \textbf{remasking} action re-masks low-confidence tokens for later refinement. 
Second, for each $i \in \mathcal{U}_t$, the token is sampled from a \textbf{temperature}-scaled categorical distribution $p_i^{(t)}(\cdot) = \mathrm{softmax}(\mathbf{l}_i^{(t)} / \tau_t)$ over raw logits $\mathbf{l}_i^{(t)}$.
Larger $\tau_t$ increases randomness, while a smaller one makes decoding greedier.
The temperature $\tau_t$ can also be \textbf{annealed} across steps to reduce entropy and stabilize the final state.

\noindent \textbf{Reward-free Guidance and Reward-Guided Search.}
Reward-free guidance, particularly classifier-free guidance (CFG)~\cite{ho2022classifier,karras2024guiding}, is a standard practice in diffusion vision models that steers sampling toward context-aligned high-likelihood regions. 
Prior work has proven that CFG can be applied to diffusion language models with a straightforward adaptation~\cite{nie2024scaling,schiff2025simple,chen2025rfg}. 
At each step $t$, the guided logits are computed as 
$\mathbf{l}^{(t)}_{\text{guided}} = \mathbf{l}^{(t)}_{\text{uncond}} + w \cdot \big( \mathbf{l}^{(t)}_{\text{cond}} - \mathbf{l}^{(t)}_{\text{uncond}} \big)$,
where $\mathbf{l}^{(t)}_{\text{cond}}$ and $\mathbf{l}^{(t)}_{\text{uncond}}$ are the conditional and unconditional logits and $w$ is the guidance scale, with $w > 1$ amplifying the conditional signal and sharpening adherence to the specified constraints.

Beyond single-trajectory guidance, search algorithms~\cite{snell2024scaling,wei2025unifying} use a global reward with two complementary views: \textit{exploring multiple trajectories in parallel}, as in Best-of-N, and \textit{evaluating the ongoing trajectory in advance}, as in soft value-based decoding (SVDD)~\cite{li2024derivative}. 
Beam search can be generalized to combine both views~\cite{didi2026scaling}, which we treat in a unified manner in Section~\ref{sec:reward}.
\section{Vanilla Sampling}
\label{sec:vanilla}

To begin our investigation of the inference-time behavior of multimodal pLMs, we revisit the default sampling of ESM3, DPLM-2, and DPLM-2.1 across four foundational tasks, with details provided in Appendix~\ref{asec:vanilla_sampling}.
Two issues motivate a closer look.
First, the inference implementations across models are not aligned: ESM3 lacks native support for stochastic unmasking, remasking, and synchronous sequence-structure sampling, which are integral to DPLM-2 and DPLM-2.1.
Second, across tasks, the four sets of defaults rely on coarse heuristics, whose rationality and optimality have not been empirically verified.
Together, these discrepancies hinder fair cross-model comparison and motivate a unified sampling framework.

To address these limitations, \textbf{we standardize inference implementations across multimodal pLMs and benchmark vanilla sampling configurations}. Specifically, we conduct a comprehensive grid search across six key dimensions: diffusion steps $T$ (spanning 1, 8, $L/8$, $L/4$, $L/2$, $L$, and optionally 100 and 500), unmasking strategies (deterministic, stochastic, or random), remasking (enabled or disabled), sampling temperature (varying in $[0, 1]$), temperature annealing (enabled or disabled), and multimodal sampling orders (synchronous, or sequence-to-structure).
The resulting optimal configurations differ considerably from the defaults across all three models and four tasks. 

\begin{figure*}[t!]
    \centering
    \includegraphics[width=\linewidth]{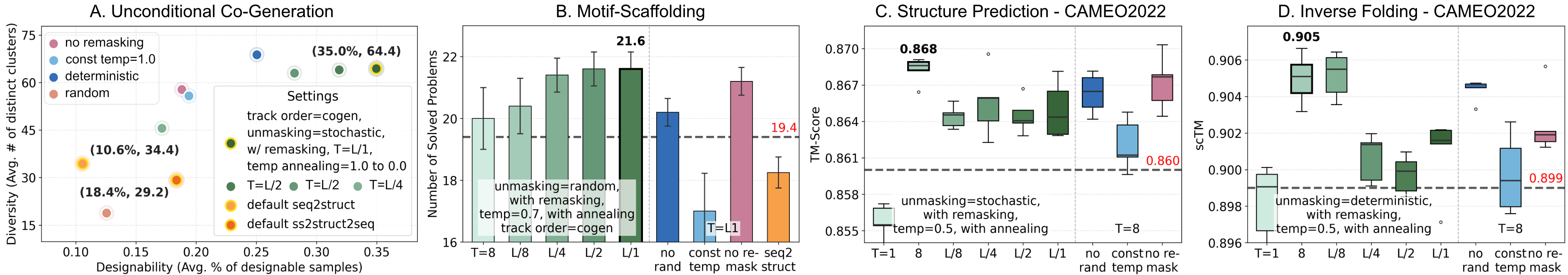}
    \captionsetup{font=small, skip=6pt}
    \caption{
    \textbf{Vanilla sampling benchmark}.
    Panels \textbf{A}-\textbf{D} show ESM3 performance on each task under the optimal vanilla sampling configuration, with ablations over total steps, per-step randomness (unmasking $\times$ temperature), temperature annealing, remasking, and multimodal sampling order.
    The optimal configuration is highlighted, and the default performance is marked for reference.
    }
    \label{fig:vanilla}
    \vspace{-12pt}
\end{figure*}

Figure~\ref{fig:vanilla} and Appendix~\ref{asec:optimal_vanilla_sampling}-\ref{asec:vanilla_sampling_investigation} present the optimal configurations and ablations for vanilla sampling settings. 
Taking ESM3 as representative, on the unconditional co-generation benchmark, enabling synchronous multimodal sampling, stochastic unmasking, and remasking \uline{unlocks ESM3 from subpar to considerably competitive}.
Under the default sequence-to-structure sampling, ESM3 produces only 52.8 designable protein samples and 34.4 unique designable clusters. With the optimized configuration, it reaches 174.8 designable samples and 64.4 clusters, on par with DPLM-2 (149.0 designable samples and 76.0 clusters).
On the motif-scaffolding benchmark, \uline{ESM3 solves obviously more problems}, increasing from an average of 19.4 to 21.6 out of 24. 
On the CAMEO2022 dataset, optimizing the sampling configuration yields \uline{consistent gains}: the average TM-score for protein structure prediction rises from 0.860 to 0.868, and the average scTM for inverse folding improves from 0.899 to 0.905. 

The optimal configurations and extent of gains differ sharply across tasks, motivating a unified, task-aware view of inference behavior through \textbf{the exploration-exploitation trade-off}. 
The trade-off should be governed by the \textit{length of the denoising trajectory} and the \textit{randomness introduced at each step} and \textit{accumulated over steps}, along which the sampling distribution is iteratively shaped. 
Our six grid-search dimensions provide concrete controls over these factors: total steps $T$ determine trajectory length; unmasking strategy and sampling temperature control per-step randomness; remasking and temperature annealing affect randomness accumulation; and the sequence-structure sampling order acts as a cross-track control in multimodal protein generation.
The corresponding exploration or exploitation tendency for each setting is summarized in Figure~\ref{fig:overview}.B.
Reading the optimized configurations through this lens, the four protein modeling tasks fall into distinct regions of the trajectory-by-randomness plane, consistent with the topology of their task-specific solution spaces.

\begin{insightbox}
\begin{itemize}[leftmargin=*,topsep=0pt,itemsep=3pt,parsep=0pt]
\item[$\dagger$] \textbf{Unconditional co-generation targets the full protein space and lies near the \uline{exploration} corner}, requiring the longest trajectories with stochastic unmasking, remasking, and high temperatures, while relying on synchronous decoding to enforce protein designability.
\item[$\dagger$] \textbf{Motif scaffolding occupies the \uline{balanced} regime}, where the solution space must simultaneously satisfy motif constraints, scaffold self-consistency, and diversity, calling for long trajectories, moderate temperature, and synchronous sequence-structure decoding.
\item[$\dagger$] \textbf{Protein structure prediction and inverse folding sit near the \uline{exploitation} corner}. The tight biophysical coupling between sequence and structure yields a narrow but non-degenerate solution space, favors short trajectories, deterministic or stochastic unmasking, and low temperatures.
\end{itemize}
\end{insightbox}

\section{Reward-Free Guidance}
\label{sec:cfg}

Classifier-free guidance (CFG)~\cite{ho2022classifier} has been widely acknowledged in general generative modeling, yet remains underexplored in multimodal pLMs.
In our investigation framework, CFG adds an orthogonal control axis to the vanilla sampling strategies. 
While vanilla sampling shapes the foundational sampling distribution, CFG instead steers the per-step logits, shifting probability mass away from the unconditional prior and toward the condition-aligned region of the solution space. 
In this section, \textbf{we design CFG strategies for four multimodal protein modeling tasks, implement them across multimodal pLMs, and examine how this logit-level control behaves across task regimes and model capabilities}.
To accommodate the dual-track nature of multimodal pLMs, we extend standard CFG to a track-wise form. 
Let $m \in \{s, z\}$ index the sequence and structure tracks, the guided logits on each track are then
{
\setlength{\abovedisplayskip}{5pt}
\setlength{\belowdisplayskip}{5pt}
\setlength{\abovedisplayshortskip}{3pt}
\setlength{\belowdisplayshortskip}{3pt}
\setlength{\jot}{3pt}
\begin{align}
    \mathbf{l}^{m,(t)}_{\text{guided}} = \mathbf{l}^{m,(t)}_{\text{uncond}} + w_m \cdot \big( \mathbf{l}^{m,(t)}_{\text{cond}} - \mathbf{l}^{m,(t)}_{\text{uncond}} \big),
\label{equ:cfg_multimodal}
\end{align}
}
where $w_m$ is the track-wise guidance scale.
Each subsection below instantiates this formulation by specifying (i) which tracks are denoised and (ii) the conditional and unconditional forward inputs that yield $\mathbf{l}^{m,(t)}_{\text{cond}}$ and $\mathbf{l}^{m,(t)}_{\text{uncond}}$.
The CFG methodology and schematic task-specific guided trajectories are also sketched in Figure~\ref{fig:overview}.C.
All experiments build on the optimal vanilla sampling strategies, with selected CFG scales reported in Appendix~\ref{asec:cfg_hyperparameters}.

\noindent \textbf{Unconditional Protein Sequence-Structure Co-Generation.}
In unconditional co-generation, multimodal pLMs receive no explicit conditioning but must maintain protein sequence-structure consistency throughout the denoising. 
Our earlier observation demonstrates that track-by-track sampling underperforms synchronous sequence-structure decoding, in which each sampling step can be viewed as a cross-modal update, with the partially decoded tokens on one track implicitly conditioning the other. 
This perspective motivates using CFG to amplify the cross-modal conditioning signal at each step. 
We thus formulate CFG as:
{
\setlength{\abovedisplayskip}{5pt}
\setlength{\belowdisplayskip}{5pt}
\setlength{\abovedisplayshortskip}{3pt}
\setlength{\belowdisplayshortskip}{3pt}
\setlength{\jot}{3pt}
\begin{align*}
    \mathbf{l}^{s,(t)}_{\text{cond}}, \mathbf{l}^{z,(t)}_{\text{cond}} &= f_\theta(\bm{s}^{(t)},\, \bm{z}^{(t)}), \\
    \mathbf{l}^{s,(t)}_{\text{uncond}} = f_\theta\big(\bm{s}^{(t)},\, \varnothing\big), &\quad
    \mathbf{l}^{z,(t)}_{\text{uncond}} = f_\theta\big(\varnothing,\, \bm{z}^{(t)}\big),
\end{align*}
}
where each unconditional logit is computed by masking the tokens on the opposite track. 
Guided logits on each track are then computed using Equation~\ref{equ:cfg_multimodal} with track-specific scales $w_s$ and $w_z$.

\begin{table*}[t]\scriptsize
\centering
\begin{threeparttable}
\definecolor{stageupfill}{HTML}{E8F5E6}
\newcommand{\stageimp}[1]{\cellcolor{stageupfill}{#1}}
\newcommand{\stagebox}[1]{\begingroup\setlength{\fboxsep}{1pt}\raisebox{0pt}[0pt][0pt]{\colorbox{stageupfill}{\strut #1}}\endgroup}
\captionsetup{font=small,skip=3pt}
\caption{Evaluation of Unconditional Sequence-Structure Co-Generation. 
Note: \stagebox{Improves} over the last investigation stage.
}
{
\begin{tabular}{l@{\hspace{3.9pt}}cccccccc}
\toprule
& & \multicolumn{2}{c}{Designable Subset} & \multicolumn{4}{c}{All Samples} \\
\cmidrule(l){3-4} \cmidrule(l){5-8}
Model & Cost* & \#Design $\uparrow$ & \#Clusters $\uparrow$ & pLDDT $\uparrow$ & 
scTM $\uparrow$ & 
\#Clusters $\uparrow$ & $\alpha$ / $\beta$ \% \\ 
\midrule
DPLM-2 $_\text{Default}$ & 2.18, 1.17 & 149.0 (126, 160)  & 76.0 (62, 84)  & 81.920 ± 8.643  & 0.906 ± 0.105  & 262.0 (248, 273)  & 38.4 / 16.9  \\ %
DPLM-2 $_\text{CFG}$ & 6.54, 3.22 & \stageimp{162.4 (157, 168)}  & \stageimp{78.4 (75, 84)}  & \stageimp{82.242 ± 8.247}  & \stageimp{0.913 ± 0.091}  & 247.4 (241, 254)  & 36.9 / \stagebox{\textbf{17.4}}  \\ %
\midrule
DPLM-2.1 $_\text{Default}$ & 2.15, 1.02 & 143.4 (129, 153)  & 86.6 (82, 97)  & 85.116 ± 7.681  & 0.904 ± 0.106  & 304.2 (295, 312)  & 43.6 / 13.6  \\ %
DPLM-2.1 $_\text{CFG}$ & 6.44, 3.00 & \stageimp{162.2 (145, 171)}  & \stageimp{96.4 (90, 104)}  & 84.995 ± 7.671  & \stageimp{0.911 ± 0.099}  & 284.8 (272, 293)  & 42.5 / \stagebox{14.2}  \\ %
\midrule
ESM3 $_\text{Default seq2struct}$ & 1.25, 1.68 & 52.8 (31, 70)  & 34.4 (24, 49)  & 61.387 ± 17.563  & 0.660 ± 0.242  & \textbf{381.4 (371, 396)}  & 48.3 / 4.3  \\ %
ESM3 $_\text{Default ss2struct2seq}$ & 3.75, 4.98 & 91.8 (90, 95)  & 29.2 (22, 45)  & 76.079 ± 13.530  & 0.762 ± 0.221  & 240.0 (229, 260)  & 71.3 / 3.5  \\ %
ESM3 $_\text{Vanilla}$ & 1.25, 1.87 & \stageimp{174.8 (167, 179)}  & \stageimp{64.4 (56, 69)}  & 68.724 ± 26.021  & 0.683 ± 0.294  & 135.0 (124, 148)  & 43.5 / \stagebox{5.3}  \\ %
ESM3 $_\text{CFG}$ & 3.75, 2.74 & \stageimp{325.0 (315, 333)} & \stageimp{116.6 (106, 123)}  & \stageimp{88.748 ± 12.007}  & \stageimp{0.931 ± 0.145}  & 194.2 (183, 202)  & 53.8 / \stagebox{10.4}  \\ %
\midrule
ESM3 $_\text{Beam - Select=Best}$ & - & \stageimp{\textbf{411.6 (406, 421)}}  & \stageimp{126.6 (122, 131)}  & \stageimp{\textbf{91.577 ± 5.208}}  & \stageimp{\textbf{0.976 ± 0.044}}  & 148.0 (142, 158)  & 45.8 / \stagebox{12.4}  \\
ESM3 $_\text{Beam - Select=Random}$ & 15.4, 11.88 & \stageimp{350.4 (336, 360)}  & \stageimp{\textbf{139.0 (132, 145)}}  & \stageimp{90.704 ± 5.525 } & \stageimp{0.964 ± 0.056}  & 177.6 (170, 186)  & 47.9 / \stagebox{11.6}  \\
ESM3 $_\text{Official SVDD}$ & - & 189.4 (175, 201)  & 88.4 (79, 94)  & 76.478 ± 15.430   & 0.881 ± 0.160  & 253.4 (241, 266)  & 51.7 / 10.4  \\
La-Proteina & - & 383.0 (370, 397) & 134.7 (125, 141)  & 83.770 ± 10.130  & 0.953 ± 0.119  & 206.0 (202, 208)  & 72.1 / 5.7  \\
\bottomrule
\end{tabular}
}
\label{tab:guided_uncon_cogen}
\begin{tablenotes}
\scriptsize
\item * denotes the computational cost at inference time, concretely the "FLOPs ($\times 10^{17}$), Runtime (Hours)".
\end{tablenotes}
\end{threeparttable}
\vspace{-10pt}
\end{table*}

As shown in Table~\ref{tab:guided_uncon_cogen}, \textbf{cross-modal CFG improves the generation designability and the diversity across all three base models}. 
For DPLM-2 and DPLM-2.1, the size of the designable subset and the number of distinct designable clusters increase moderately over their default sampling. 
On ESM3, the gains are much more pronounced. 
Across all generated samples, ESM3 with CFG attains better sequence foldability (average pLDDT of 88.748) and stronger sequence-structure self-consistency (average scTM of 0.931). 
The number of designable protein samples reaches 325.0, nearly doubling that of the optimal vanilla sampling (174.8), while the number of distinct designable clusters rises to 116.6, about 1.8 times that of the optimal vanilla sampling (64.4). 
In prior protein co-design studies, ESM3 has commonly been regarded as an underperforming baseline. 
Our results show that \textbf{once equipped with optimal vanilla sampling and a properly designed CFG, ESM3 delivers a strong performance}, surpassing both DPLM-2 and DPLM-2.1 by a clear margin.

\begin{figure}[t]
    \centering
    \includegraphics[width=\linewidth]{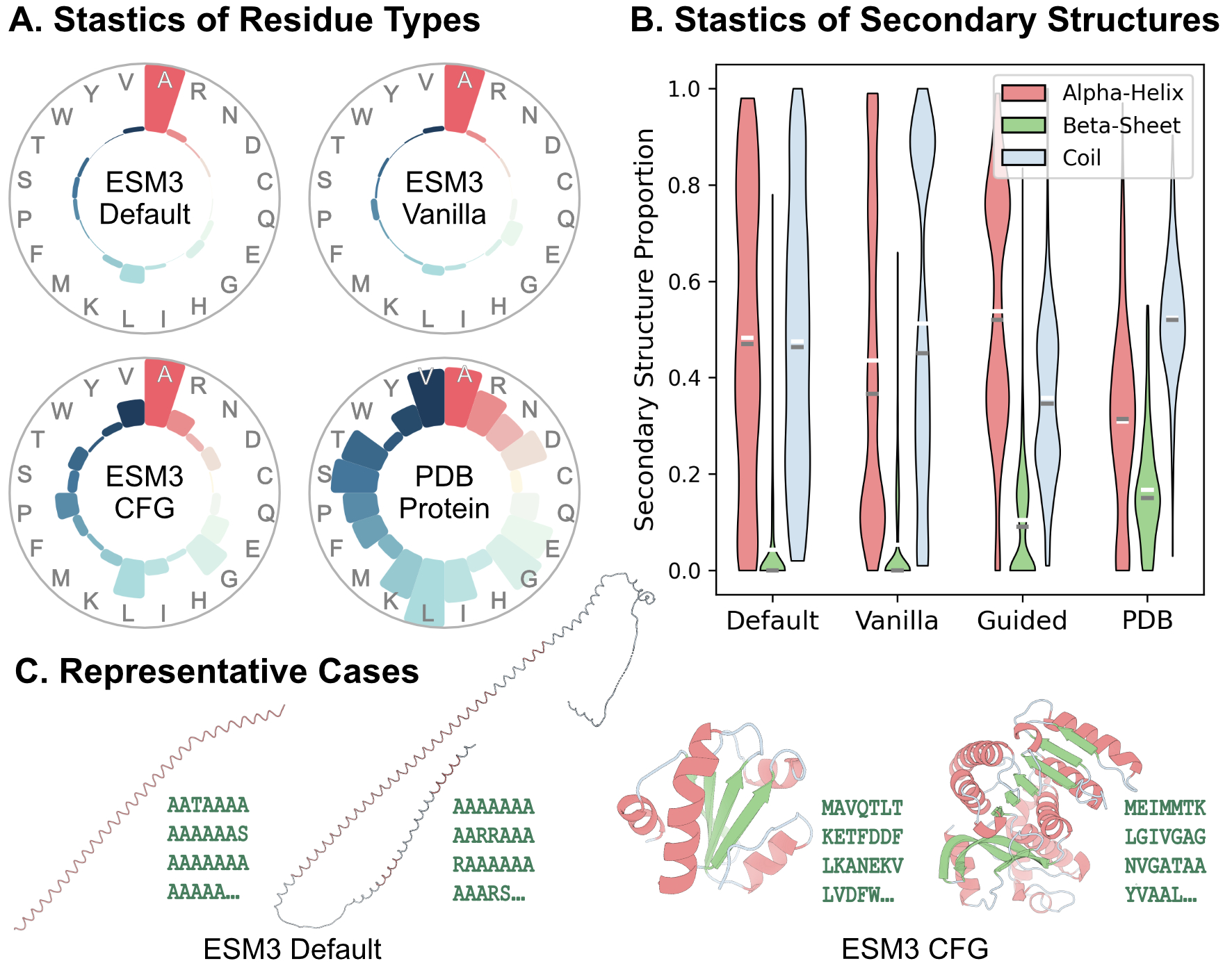}
    \captionsetup{font=small,skip=6pt}
    \caption{\textbf{Qualitative Analysis.}
    \textbf{A-B.} Optimal vanilla sampling and CFG alleviate the abnormal properties of ESM3-generated proteins. 
    \textbf{C.} Visualizations of representative cases.
    }
    \label{fig:property}
    \vspace{-12pt}
\end{figure}

Beyond the quantitative gains, the qualitative behavior of ESM3 also changes.
Under default sampling, ESM3 often produces implausible proteins with repetitive residues and a biased secondary-structure composition.
Then, \textbf{cross-modal CFG can make the generations more "protein-like".}
As shown in Figure~\ref{fig:property}, applying our cross-modal CFG reduces residue repetition and brings the $\alpha$-helix/$\beta$-sheet/coil proportions closer to those of natural proteins. 
In addition, Appendix~\ref{asec:pdb_tm_uncon_cogen} presents a PDB alignment analysis that further demonstrates that cross-modal CFG makes the generations structurally more similar to PDB proteins. 
We interpret this qualitative effect as an implicit regularization of joint multimodal sampling. 
By up-weighting cross-track conditional logits, CFG discourages the tendency for one track to move toward plausible states in isolation while having low joint probability with the other track.
Prior work sought to mitigate the repetition-induced generation collapse in pLMs by injecting external steering vectors~\cite{zhang2026controlling}, yet our results show that internal cross-modal steering can already be highly effective.

\noindent \textbf{Motif-Scaffolding.}
In motif-scaffolding, multimodal pLMs sample both sequence and structure tracks to generate coherent protein scaffolds for specific motifs. 
Success requires both preserving local motifs and maintaining global protein designability.
The former emphasizes the explicit conditioning of the motifs. 
The latter, along with our observation of the superiority of synchronous decoding over sequence-to-structure sampling, highlights the implicit cross-modal conditioning. 
We therefore use a track-wise CFG that emphasizes both motif and cross-modal alignments.
The conditional logits on both tracks are obtained from
$\mathbf{l}^{s,(t)}_{\text{cond}}, \mathbf{l}^{z,(t)}_{\text{cond}} = f_\theta(\bm{s}^{(t)},\, \bm{z}^{(t)})$. 
The unconditional logits are computed by removing motif and cross-modal information from each track:
{
\setlength{\abovedisplayskip}{5pt}
\setlength{\belowdisplayskip}{5pt}
\setlength{\abovedisplayshortskip}{3pt}
\setlength{\belowdisplayshortskip}{3pt}
\setlength{\jot}{3pt}
\begin{align*}
    \mathbf{l}^{s,(t)}_{\text{uncond}} = f_\theta\big(\bm{s}^{(t)}\setminus\bm{s}_{\text{motif}},\, \varnothing\big), \\
    \mathbf{l}^{z,(t)}_{\text{uncond}} = f_\theta\big(\varnothing,\, \bm{z}^{(t)}\setminus\bm{z}_{\text{motif}}\big). 
\end{align*}
}
The guided logits on each track then follow Equation~\ref{equ:cfg_multimodal} with track-specific scales $w_s$ and $w_z$.
\begin{table}\scriptsize
\centering
\begin{threeparttable}
\captionsetup{font=small,skip=3pt}
\caption{Evaluation of Motif-Scaffolding.}
{
\definecolor{stageupfill}{HTML}{E8F5E6}
\newcommand{\stageimp}[1]{\cellcolor{stageupfill}{#1}}
\setlength{\tabcolsep}{2.3pt}
\begin{tabular}{l@{\hspace{0pt}}cccc}
\toprule
Model & Cost* & \#Solved $\uparrow$ & Succ. Rate $\uparrow$ & \#Clusters $\uparrow$ \\
\midrule
DPLM-2 $_\text{Default}$ & 2.88, 1.37 & 15.6 (15, 16)  & 15.3 ± 0.3 \%  & 72.6 (63, 83)  \\
DPLM-2 $_\text{Vanilla}$ & 0.47, 0.42 & \stageimp{16.8 (16, 17)}  & \stageimp{20.6 ± 0.5 \%}  & \stageimp{72.8 (68, 80)}  \\
DPLM-2 $_\text{CFG}$ & 1.41, 0.75 & 15.8 (15, 17)  & 20.4 ± 0.2\%  & 71.6 (63, 81)  \\
\midrule
DPLM-2.1 $_\text{Default}$ & 2.83, 1.15 & 14.2 (14, 15)  & 18.7 ± 0.6 \%  & 51.0 (45, 55)  \\
DPLM-2.1 $_\text{Vanilla}$ & 2.83, 1.15 & \stageimp{15.4 (15, 16)}  & \stageimp{19.4 ± 0.6 \%}  & \stageimp{66.8 (61, 72)}  \\
DPLM-2.1 $_\text{CFG}$ & 8.49, 3.33 & 15.0 (14, 16)  & 19.3 ± 0.5 \%  & 61.2 (56, 68)  \\
\midrule
ESM3 $_\text{Default}$ & 0.18, 1.11 & 19.4 (19, 20)  & 19.6 ± 0.8 \%  & 94.2 (85, 98)  \\
ESM3 $_\text{Vanilla}$ & 0.28, 1.66 & \stageimp{21.6 (21, 22)}  & \stageimp{30.9 ± 1.1 \%}  & \stageimp{110.0 (100, 124)}  \\
ESM3 $_\text{CFG}$ & 0.84, 1.67 & \stageimp{22.8 (22, 23)}  & \stageimp{37.5 ± 1.0 \%}  & \stageimp{179.4 (165, 198)}  \\
\midrule
ESM3 $_\text{Beam}$ & 3.65, 15.64 & \stageimp{\textbf{23.0 (23, 23)}}  & \stageimp{\textbf{49.9 ± 0.7 \%}}  & \stageimp{\textbf{261.0 (250, 278)}}  \\
La-Proteina $_\text{scaff}$ & - & 22.0 (22, 22)  & 42.7 ± 0.7 \%  & 256.6 (247, 272)  \\
\bottomrule
\end{tabular}
}
\label{tab:guided_scaffolding}
\begin{tablenotes}
\scriptsize
\item * Inference cost in "FLOPs ($\times 10^{17}$), Runtime (Hours)".
\end{tablenotes}
\end{threeparttable}
\vspace{-12pt}
\end{table}

Table~\ref{tab:guided_scaffolding} summarizes the results.
\textbf{While CFG brings no further gains over optimal vanilla sampling for DPLM-2 and DPLM-2.1, it substantially improves ESM3}.
In 4 out of 5 independent runs, ESM3 solves 23 of the 24 benchmark cases, outperforming DPLM-2/-2.1, and ESM3 itself without guidance. 
The average success rate rises to 37.5\%, nearly doubling that of the default sampling and about 1.2 times that of the optimal vanilla sampling. 
The diversity of successful scaffolds is improved accordingly: ESM3 with CFG produces 179.4 unique solution clusters on average, nearly twice that of the default sampling (94.2) and about 1.6 times that of the optimal vanilla sampling (110.0). 
We interpret this contrast as a difference in \textbf{how reliably each base model encodes the motif and cross-modal conditions}. 
ESM3 appears to encode these conditions strongly enough for effective logit-level steering, whereas DPLM-2 and DPLM-2.1 do not.
We further provide an ablation on the dual-condition CFG design in Appendix~\ref{asec:scaffolding_cfg_ablation}. 

\begin{table*}[t]\scriptsize
\centering
\begin{threeparttable}
\captionsetup{font=small,skip=2.3pt}
\caption{Evaluation of Protein Structure Prediction (Left) and Inverse Folding (Right).}
\label{tab:guided_folding}
\label{tab:guided_inv_folding}
{
\definecolor{stageupfill}{HTML}{E8F5E6}
\newcommand{\stageimp}[1]{\cellcolor{stageupfill}{#1}}
\setlength{\tabcolsep}{2pt}
\begin{tabular}{l@{\hspace{1pt}}ccccc@{\hspace{6pt}}c||@{\hspace{6pt}}l@{\hspace{1pt}}ccccc}
\toprule
~ & \multicolumn{3}{c}{CAMEO} & \multicolumn{2}{c}{PDB Date Split} &  & ~ & \multicolumn{3}{c}{CAMEO} & \multicolumn{2}{c}{PDB Date Split} \\
\cmidrule(lr){2-4} \cmidrule(lr){5-6} \cmidrule(lr){9-11} \cmidrule(lr){12-13}
Model & Cost* & RMSD $ \downarrow $ & TM-score $ \uparrow $ & RMSD $ \downarrow $ & TM-score $ \uparrow $ &  & Model & Cost*  & scTM $ \uparrow $ & AAR & scTM $ \uparrow $ & AAR \\
\midrule
DPLM-2 $_\text{Default}$ & 18.5, 6.2 & 7.483±6.126  & 0.786±0.170  & 5.253±5.143  & 0.836±0.144  &  & DPLM-2 $_\text{Default}$ & 18.5, 6.2 & 0.870±0.158 & 48.1 & 0.912±0.111  & 55.6 \\
DPLM-2 $_\text{Vanilla}$ & 5.6, 2.8 & 7.513±6.178  & \stageimp{0.790±0.167}  & \stageimp{5.249±5.103}  & \stageimp{0.837±0.141}  &  & DPLM-2 $_\text{Vanilla}$ & 0.2, 0.1 & \stageimp{0.875±0.153} & 48.2 & \stageimp{0.917±0.111}  & 55.6 \\
DPLM-2 $_\text{CFG}$ & 11.2, 4.1 & \stageimp{7.414±6.112} & 0.786±0.169  & \stageimp{5.178±5.029}  & 0.835±0.142  &  & DPLM-2 $_\text{CFG}$ & 0.4, 0.1 & \stageimp{0.877±0.151}  & 47.1 & \stageimp{0.922±0.101}  & 54.9 \\
\midrule
DPLM-2.1 $_\text{Default}$ & 18.2, 5.5 & 6.297±6.181 & 0.822±0.170  & -  & -  &  & DPLM-2.1 $_\text{Default}$ & 18.2, 5.5 & 0.875±0.144  & 53.3 &  -  & - \\
DPLM-2.1 $_\text{Vanilla}$ & 1.5, 0.5 & \stageimp{6.259±6.051}  & \stageimp{0.824±0.167}  &  -  & -  &  & DPLM-2.1 $_\text{Vanilla}$ & 1.5, 0.5 & \stageimp{0.879±0.142}  & 53.4 &  -  & - \\
DPLM-2.1 $_\text{CFG}$ & 2.9, 0.9 & \stageimp{6.104±5.809}  & \stageimp{0.824±0.167}  &  -  & -  &  & DPLM-2.1 $_\text{CFG}$ & 2.9, 0.9 & \stageimp{0.880±0.148}  & 52.1 &  -  & - \\
\midrule
ESM3 $_\text{Default}$ & 0.1, 0.3 & 5.334±6.280  & 0.860±0.168  & 4.053±4.867  & 0.882±0.152  &  & ESM3 $_\text{Default}$ & 0.8, 1.4 & 0.899±0.143  & 46.4 & 0.944±0.083  & 49.6 \\
ESM3 $_\text{Vanilla}$ & 0.8, 1.2 & 5.563±7.108  & \stageimp{0.868±0.161}  & 4.872±7.178  & \stageimp{0.885±0.150}  &  & ESM3 $_\text{Vanilla}$ & 0.8, 1.4 & \stageimp{0.905±0.134}  & 48.4 & \stageimp{0.947±0.085}  & 52.0  \\
ESM3 $_\text{CFG}$ & 1.6, 1.3 & \stageimp{5.144±6.129}  & \stageimp{0.870±0.157}  & \stageimp{3.969±4.823}  & \stageimp{0.889±0.142}  &  & ESM3 $_\text{CFG}$ & 1.6, 1.5 & \stageimp{0.910±0.135} & 45.7 & \stageimp{0.954±0.075}  & 49.1 \\
\midrule
ESM3 $_\text{Beam}$ & 35.6, 36.2 & 5.221±6.565  & \stageimp{0.872±0.157}  & \stageimp{3.744±4.542}  & \stageimp{0.893±0.140}  &  & ESM3 $_\text{Beam}$ & 47.4, 50.0 & \stageimp{\textbf{0.914±0.129}}  & 46.0 & \stageimp{\textbf{0.956±0.069}}  & 49.3 \\
ESMFold & - & \textbf{3.961±4.795} & \textbf{0.898±0.147}  & \textbf{2.661±3.916}  & \textbf{0.931±0.112}  &  & ProteinMPNN & - & 0.902±0.145  & 42.4 & 0.953±0.075  & 46.8 \\
\bottomrule
\end{tabular}
}
\begin{tablenotes}
\scriptsize
\item * Inference cost in "FLOPs ($\times 10^{15}$), Runtime (Minutes)".
\end{tablenotes}
\end{threeparttable}
\vspace{-12pt}
\end{table*}

\noindent \textbf{Protein Structure Prediction.}
For structure prediction, multimodal pLMs condition on the input sequence $\bm{s}$ and perform iterative discrete diffusion sampling along the structure track $\bm{z}$. 
At each denoising step $t$, the conditional structure logits are obtained through a forward pass $\mathbf{l}^{z,(t)}_{\text{cond}} = f_\theta(\bm{s},\, \bm{z}^{(t)})$. 
The corresponding unconditional logits are obtained by fully masking the sequence track and executing an additional forward pass, yielding $\mathbf{l}^{z,(t)}_{\text{uncond}} = f_\theta(\varnothing,\, \bm{z}^{(t)})$. 
Guided sampling is then performed by extrapolating the two via Equation~\ref{equ:cfg_multimodal}.
Table~\ref{tab:guided_folding} summarizes the evaluation results. 

For DPLM-2 and DPLM-2.1, CFG yields modest RMSD improvements and no consistent TM-score gains. 
In contrast, \textbf{ESM3 benefits clearly from CFG across both datasets and metrics}, achieving an average RMSD of 5.144 and TM-score of 0.870 on CAMEO2022, and an RMSD of 3.969 with TM-score of 0.889 on PDB Date.
We attribute this contrast to \textbf{the narrow solution space in structure prediction and the differing sequence-conditional capacity across base models}. 
As a given sequence admits only a small set of plausible folds, CFG should be useful when a pLM has not yet fully exploited its sequence condition.
DPLM-2 and DPLM-2.1 already operate near their conditional capacity under optimal vanilla sampling,
whereas ESM3, with the strongest capability of the three, leaves more room above its conditional optimum for logit-level steering.
Additionally, a divergence between RMSD and TM-score performance in ESM3 is analyzed in Appendix~\ref{asec:detailed_explain}.

\noindent \textbf{Inverse Folding.}
For inverse folding, multimodal pLMs iteratively sample along the sequence track $\bm{s}$, conditioned on the target structure $\bm{z}$. 
We formulate CFG for this task symmetrically to structure prediction.
At each denoising step $t$, the conditional sequence logits are $\mathbf{l}^{s,(t)}_{\text{cond}} = f_\theta(\bm{s}^{(t)},\, \bm{z})$. 
The corresponding unconditional logits are obtained by dropping the structural condition and executing an additional forward pass, yielding $\mathbf{l}^{s,(t)}_{\text{uncond}} = f_\theta(\bm{s}^{(t)},\, \varnothing)$.

Table~\ref{tab:guided_inv_folding} shows that \textbf{CFG consistently improves structural consistency and reduces sequence similarity to the native reference across base models}. 
For ESM3, the average scTM reaches 0.910 on CAMEO2022 and 0.954 on PDB Date, while the AAR declines slightly.
We interpret this decline in light of the one-to-many nature of inverse folding and the exploitation effect of CFG. 
It improves structural consistency by shifting pLM probability mass toward protein sequences that better satisfy the target structure, without requiring them to recover the native reference.

\noindent \textbf{Inference Efficiency.} 
Alongside benchmark performance, Tables~\ref{tab:guided_uncon_cogen}-\ref{tab:guided_inv_folding} also report the computational costs across inference protocols, revealing the inefficiency of default sampling and the superior performance-efficiency balance of our CFG strategies. 
See Appendix~\ref{asec:effciency_analysis} for detailed discussions.

\begin{insightbox}
\begin{itemize}[leftmargin=*,topsep=0pt,itemsep=3pt,parsep=0pt]
\item[$\dagger$] \textbf{CFG provides a simple and general control for multimodal protein modeling}. By casting diverse tasks into conditional generation and dropping the appropriate conditions in the unconditional branch, it consistently improves the protein generation performance.
\item[$\dagger$] \textbf{CFG helps more when the task condition leaves a larger solution space, giving more room to steer the sampling}. It is evidenced by the larger gains on unconditional cogeneration and motif-scaffolding tasks, which have a considerable exploratory nature.
\item[$\dagger$] \textbf{The benefit of CFG is model-dependent and scales with how well the base model has internalized the corresponding conditional signal.} Stronger multimodal pLMs expose more usable condition-aligned logit changes for steering, tend to gain more from CFG.
\end{itemize}
\end{insightbox}

\section{Reward-Guided Search}
\label{sec:reward}

Our study of vanilla sampling and reward-free guidance shows that the performance of multimodal pLMs can be improved by optimizing the sampling trajectory.
However, they still generate only a single trajectory per run and rely on per-step token-level logits, which assess local probabilities rather than the global quality of the final output, leaving potentially superior trajectories unexplored.
To address this limitation, 
\textbf{we lift control from individual steps to entire trajectories: multiple trajectories are maintained in parallel and selected using global rewards.}
Notably, this section focuses exclusively on ESM3. 
Our earlier results show that it is the only base model with consistent room for logit-level steering, making it the natural base for studying trajectory-level search.
Implementation details are presented in Appendix~\ref{asec:rewarded_search}.

Specifically, we adapt beam search for the inference of multimodal pLMs, as illustrated in Figure~\ref{fig:overview}.D and Algorithm~\ref{alg:beam}, and as explained in detail in Appendix~\ref{asec:beam_search_multimodal_plms}. 
Built on top of the underlying sampler, \uline{the search layer is decoupled from vanilla sampling and CFG choices}.

Let $\mathcal{T}^{(t)}$ denote the retained sampling trajectories at step $t$, where each trajectory is a partially denoised protein $(\bm{s}^{(t)}, \bm{z}^{(t)})$. 
Starting from $\mathcal{T}^{(T)}$ with beam width $N$, branching factor $B$, and scoring interval $K$, we perform expand-score-select operations periodically:
{
\setlength{\abovedisplayskip}{5pt}
\setlength{\belowdisplayskip}{5pt}
\setlength{\abovedisplayshortskip}{3pt}
\setlength{\belowdisplayshortskip}{3pt}
\setlength{\jot}{3pt}
\begin{align*}
	\mathcal{C}^{(t)} &\gets \textsc{Expand}(\mathcal{T}^{(t)}, B), \quad \text{if } t \bmod K = 0, \\
	\hat r(\bm{c}) &\gets r(\textsc{QuickUnmask}(\bm{c})), \quad \bm{c} \in \mathcal{C}^{(t)}, \\
	\mathcal{T}^{(t-1)} &\gets \textsc{Select}(\mathcal{C}^{(t)}, \hat r, N).
\end{align*}
}
Every $K$ steps (when $t \bmod K = 0$), the search maintains $N$ parallel trajectories and expands each trajectory into $B$ candidates via independent one-step samplings. 
While these candidates are still partially masked, we use the estimated reward $\hat r$: each candidate is denoised by one-pass $\textsc{QuickUnmask}(\cdot)$ and then scored by a reward $r(\cdot)$. The search then prunes the beam set back to width $N$ under $\hat r$ via a selection rule $\textsc{Select}(\cdot)$. 

At non-expansion steps (when $t \bmod K \neq 0$), all retained trajectories are simply advancing.
After the final step, a single output is selected from the beam set $\mathcal{T}^{(0)}$ using $\textsc{Select}(\cdot)$ again.

Within this framework, task-specific design reduces to the reward function $r(\cdot)$ and the selection rule $\textsc{Select}(\cdot)$. 
To preserve the ``all-in-one'' nature of multimodal pLMs and keep comparisons with vanilla sampling and CFG fair, we use model-internal signals as rewards, i.e., pTM scores produced during protein structure de-tokenization.
The structure track is scored directly by structural pTM, whereas the sequence track is first greedily folded into structure tokens and then scored, named foldability pTM. 
Then, the selection rule should match each task's exploration-exploitation profile: for protein structure prediction, inverse folding, and motif-scaffolding, $\textsc{Select}(\cdot)$ returns the top candidates ranked by structural pTM, foldability pTM, or their sum, respectively. 
For unconditional co-generation, we instead randomly select candidates that satisfy structural and foldability constraints (pTM $> 0.8$ on both tracks), thereby preserving diversity while enforcing a quality threshold. 

Tables~\ref{tab:guided_uncon_cogen}-\ref{tab:guided_inv_folding} also report the performance of ESM3 with reward-guided search, alongside specialist baselines. 
\textbf{Across all four tasks, beam search consistently improves over the CFG stage and enables a single foundation protein model to match or surpass task-specific SOTA systems without task-specific training or external rewards.} 
For unconditional co-generation, ESM3 attains the strongest designability across all baselines. The threshold-based random selection variant raises the number of unique designable clusters from 116.6 to 139.0, surpassing La-Proteina~\cite{geffner2025laproteina}.
We also include the official SVDD implementation in ESM3, which works atop the default sequence-to-structure sampling. Its limited performance further suggests that our improvements from vanilla sampling and CFG are complementary to reward-guided search. 
On motif-scaffolding, ESM3 solves 23 of 24 problems in every run, raising the success rate from 37.5\% to 53.3\% and the number of unique clusters from 179.4 to 245.2, exceeding La-Proteina's motif-scaffolding variant. 
For protein structure prediction, ESM3 further reduces the average RMSD and lifts the TM-score on PDB Date, narrowing the gap to ESMFold~\cite{lin2023esm2}.
For inverse folding, ESM3 achieves the best scTM on both CAMEO2022 and PDB Date (0.914 and 0.956), surpassing ProteinMPNN~\cite{dauparas2022robust}. 

Notably, beam search delivers its full benefit when the underlying vanilla sampling is tuned to be slightly more exploratory than its single-trajectory optimum, as leaving room for reward-driven selection to discriminate among candidate trajectories.  
Details are deferred to Appendix~\ref{asec:explore_in_beam}.
Besides, as transparently presented in Table~\ref{tab:guided_uncon_cogen}-\ref{tab:guided_inv_folding}, beam search incurs considerable computational overhead at inference to achieve improved benchmark performance. Computational efficiency issues are discussed in detail in Appendix~\ref{asec:effciency_analysis}.

\begin{insightbox}
\begin{itemize}[leftmargin=*,topsep=0pt,itemsep=3pt,parsep=0pt]
\item[$\dagger$] \textbf{Multimodal pLMs can support a simple and effective internal loop at inference time.} Searching across parallel trajectories with model-internal rewards further improves protein modeling performance without additional training or external judges.
\item[$\dagger$] \textbf{Our three-stage evaluation pushes multimodal pLMs to the level of task-specific SOTA systems across multiple protein modeling tasks}. This may reshape the community's view of multimodal pLMs' position, which has so far been formed by their performance under default sampling.
\end{itemize}
\end{insightbox}
\section{Conclusion}

In this paper, we elucidate the inference space of multimodal pLMs.
Across three base models on four tasks, \textbf{vanilla sampling}, \textbf{reward-free guidance}, and \textbf{reward-guided search} act in complementary ways and together yield consistent, clear-margin gains over each model's default configuration, bringing multimodal pLMs close to, or even surpass, task-specific specialist models. 
Two unifying patterns emerge along the way: 
(1) inference preferences are largely \uline{task-oriented}; and (2) the exploration-exploitation trade-off can be navigated \uline{bottom-up}---first centering the foundational sampling \uline{distribution}, then steering its \uline{logits} toward the condition, and finally selecting among \uline{trajectories} with a global quality signal. 

\section*{Limitations}
\label{asec:limitations}

This work studies inference-time strategies for fixed multimodal pLMs on controlled benchmark tasks. 
Our goal is to improve the final-sample quality through vanilla sampling, classifier-free guidance, and reward-guided search, using the model's internal logits and quality signals. 
Although these results are useful across the tasks studied here, they do not directly translate to broader notions of biological utility or experimental success. 
We do not study richer deployment settings in which multimodal pLMs interact with external predictive models, human experts, laboratory feedback, or multi-stage protein design pipelines. 
These remain important directions for future work.

Besides, our analysis of performance-efficiency trade-offs remains at a very basic stage.
Since the current work does not take computational efficiency as the optimization objective, the reported FLOPs and runtime objectively reflect the computational efficiency of inference protocols optimized purely for performance. We consider the multi-objective optimization of inference efficiency and benchmark performance under limited computational resources as future work.

\section*{Ethical Considerations}

The primary societal impact of this work is to support scientific research and beneficial clinical and biomedical applications. 
However, as with any generative technology in biology, there is a theoretical risk of misuse in designing harmful biomolecules, such as pathogenic proteins. 
We believe this risk is mitigated by the fact that our study investigates the inference procedures of models rather than introducing new biological capabilities, operates on established computational benchmarks, and does not autonomously propose or experimentally validate hazardous biological designs. 
All data used in this work were obtained from publicly available sources, and no ethical approval was required.

\bibliography{references}  

\appendix

\clearpage
\section{Appendix}

\paragraph{Table of Appendix:}
\begin{itemize}[leftmargin=*,itemsep=3pt,parsep=0pt]
    \item \ref{asec:mplm_preliminary}. Supplementary Preliminaries
    \begin{itemize}[leftmargin=*,itemsep=3pt,parsep=0pt]
        \item \ref{asec:tokenization}. Protein Structure Tokenization
        \item \ref{asec:base_model_descriptions}. Base Model Descriptions
        \item \ref{asec:evaluation_pipelines}. Evaluation Pipelines
    \end{itemize}
    \item \ref{asec:vanilla_sampling}. Details in Vanilla Sampling
    \begin{itemize}[leftmargin=*,itemsep=3pt,parsep=0pt]
        \item \ref{asec:default_sampling}. Default Sampling Strategies
        \item \ref{asec:optimal_vanilla_sampling}. Optimal Vanilla Sampling Strategies
        \item \ref{asec:vanilla_sampling_investigation}. Ablation Study
    \end{itemize}
    \item \ref{asec:cfg}. Details in Reward-free Guidance
    \begin{itemize}[leftmargin=*,itemsep=3pt,parsep=0pt]
        \item \ref{asec:cfg_hyperparameters}. Hyperparameter Selection
        \item \ref{asec:pdb_tm_uncon_cogen}. PDB Alignment Analysis on Unconditional Co-Generation Proteins.
        \item \ref{asec:scaffolding_cfg_ablation}. Ablation Study on Motif-Scaffolding CFG
        \item \ref{asec:detailed_explain}. A Detailed Protein Structure Prediction Observation
    \end{itemize}
    \item \ref{asec:rewarded_search}. Details in Reward-guided Search
    \begin{itemize}[leftmargin=*,itemsep=3pt,parsep=0pt]
        \item \ref{asec:rewarded_impl}. Implementation of Specialist Models
        \item \ref{asec:beam_search_multimodal_plms}. Beam Search for Multimodal pLMs
        \item \ref{asec:rewarded_hyperparameters}. Hyperparameter Selection
        \item \ref{asec:explore_in_beam}. Beam Search: Explore-then-Select
    \end{itemize}
    \item \ref{asec:effciency_analysis}. Inference Efficiency Analysis
\end{itemize}

\subsection{Supplementary Preliminaries}
\label{asec:mplm_preliminary}

\subsubsection{Protein Structure Tokenization}
\label{asec:tokenization}

For a protein with $ L $ residues, its sequence is formulated as $ \bm{s} = \left( s_1, s_2, \dots, s_L \right) $, where each $ s_i \left( 1 \leq i \leq L \right) $ is a categorical variable denotes the identity of the $i$-th residue, generally involved in 20 standard amino acids $ \mathbb{S}^{20} = \{\texttt{A}, \texttt{R}, \dots, \texttt{V}\} $.
Meanwhile, the protein structure is firstly formulated by $ \bm{x} = \left( x_1, x_2, \dots, x_L \right) $, where $ x_i \in \mathbb{R}^{n_i \times 3} $ is the Cartesian coordinates of the $i$-th residue's atoms.

To accommodate the discrete nature of language models, multimodal pLMs~\cite{hayes2025simulating,wang2024dplm,hsieh2025elucidating} generally handle protein structures using quantization-based tokenizers. 
Similar to image tokenization, the protein structure tokenization process can be summarized under a VQ-VAE~\cite{van2017neural} framework:
{
\setlength{\abovedisplayskip}{5pt}
\setlength{\belowdisplayskip}{5pt}
\setlength{\abovedisplayshortskip}{3pt}
\setlength{\belowdisplayshortskip}{3pt}
\setlength{\jot}{3pt}
\begin{align*}
\bm{x} \xrightarrow{\text{encoder \& quantizer}} \bm{z} \xrightarrow{\text{decoder}} \hat{\bm{x}},
\end{align*}
}
where an encoder and a quantizer convert protein structure coordinates $\bm{x}$ into $L$ discrete tokens $\bm{z} = (z_1, \dots, z_L)$ each within a finite-size codebook, and the decoder reconstructs 3D coordinates $\hat{\bm{x}}$. 
Notably, following the practice of AlphaFold~\cite{jumper2021highly} and ESMFold~\cite{lin2023esm2}, the decoder of ESM3's structure tokenizer includes dedicated heads to compute structural quality scores, typically the predicted template modeling (pTM) score.

\subsubsection{Base Model Descriptions}
\label{asec:base_model_descriptions}

\textbf{DPLM-2}~\cite{wang2024dplm} is a multimodal protein language model that extends a sequence-only pLM to model both sequence and structure. 
To incorporate structural information, it converts 3D coordinates $\bm{x}$ into discrete tokens $\bm{z}$ using a lookup-free quantization tokenizer\footnote{https://huggingface.co/airkingbd/struct\_tokenizer}.
Trained on high-quality protein data, DPLM-2 learns the joint distribution of protein sequence and structure, together with their marginals and conditionals. 
In our implementation, we use the pretrained 650M checkpoint\footnote{https://huggingface.co/airkingbd/dplm2\_650m}, following the official instructions\footnote{https://github.com/bytedance/dplm}.

\noindent
\textbf{DPLM-2.1}~\cite{hsieh2025elucidating} likewise models the joint distribution of protein sequence and structure. 
For structural modeling, it uses the same LFQ tokenizer as DPLM-2, with a vocabulary size of $2^{13}=8192$, but predicts the 13 binary bits of each token rather than the corresponding 8192-way index. 
This finer-grained supervision should improve protein structure modeling.
In our implementation, we use the pretrained 650M checkpoint\footnote{https://huggingface.co/airkingbd/dplm2\_bit\_650m}, following the official instructions\footnote{https://github.com/bytedance/dplm}.

\noindent
\textbf{ESM3}~\cite{hayes2025simulating} is a multimodal generative language model that reasons over the sequence, structure, and function of proteins.
Besides tokenized sequence and structure tracks, it also supports four additional parallel tracks, such as secondary structure labels, and can take structure coordinates as input.
In this work, we focus on the sequence and structure tracks and enable coordinate input to improve protein structure modeling.
In our implementation, we use the open-source ESM3-Open (1.4B) checkpoint\footnote{https://huggingface.co/EvolutionaryScale/esm3-sm-open-v1}, following the official codebase\footnote{https://github.com/evolutionaryscale/esm}.

For notational uniformity, we denote the protein structure by discrete tokens $\bm{z}$ throughout the main text. In practice, however, we follow each model's recommended input format: ESM3 takes raw backbone coordinates $\bm{x}$ and motif coordinates $\bm{x}_{\text{motif}}$ for inverse folding and motif scaffolding, whereas DPLM-2 and DPLM-2.1 require the tokenized forms $\bm{z}$ and $\bm{z}_{\text{motif}}$. 
This input choice does not affect the generation formulation in Equation~\ref{equ:cfg_multimodal}.

In our practice, all experiments were conducted on a single NVIDIA H20 (96GB) GPU. 
To compare the computational efficiency of different inference protocols, we also report the floating-point operations per second (FLOPs) and wall-clock runtime for each experiment.

\subsubsection{Evaluation Pipelines}
\label{asec:evaluation_pipelines}

\textbf{Unconditional protein sequence-structure co-generation} jointly produces protein sequence and structure with only chain length specified~\cite{geffner2025laproteina,zhou2025hd}. Following established protocols~\cite{wang2024dplm,geffner2025laproteina}, we sample 100 proteins at each target length (100, 200, 300, 400, and 500 residues), and report designability and diversity over both the designable subset and the full sample set. For each generated sample, we refold the sequence with ESMFold~\cite{lin2023esm2}, where pLDDT is ESMFold's prediction confidence, and scRMSD and scTM measure the consistency between the ESMFold-predicted and the generated structures. A sample is classified as designable if $\text{scRMSD} < 2.0~\text{\AA}$. On the \uline{designable subset}, designability is summarized by the number of designable samples and diversity by the number of Foldseek~\cite{van2024fast} clusters of these samples. On the \uline{full sample set}, designability is summarized by pLDDT and scTM, and diversity by the number of Foldseek clusters of all generated samples. 
Additionally, the secondary structure proportions are also calculated.

\textbf{Motif scaffolding} aims to generate protein scaffold structures that correctly embed specified target motifs~\cite{geffner2025laproteina,watson2023novo}. Following established protocols~\cite{wang2024dplm,yim2024improved}, we generate 100 candidate scaffolds for each of the 24 benchmark problems, with scaffold length and motif order determined according to the specifications. For each generated candidate, we refold the sequence with ESMFold~\cite{lin2023esm2} and obtain scRMSD between the ESMFold-predicted structure and the generated structure, while motif-RMSD is computed directly between the motif residues of the generated structure and those of the ground-truth motif. A candidate is counted as successful if and only if it satisfies both global designability ($\text{scRMSD} < 2.0~\text{\AA}$) and local motif preservation ($\text{motif-RMSD} < 1.0~\text{\AA}$)~\citep{zheng2025motifbench}. The diversity of successful designs is quantified via Foldseek clustering~\cite{van2024fast}.

\textbf{Protein structure prediction} infers a protein's 3D structure from its amino acid sequence~\citep{jumper2021highly,lin2023esm2}. Following established protocols~\cite{wang2024dplm,hsieh2025elucidating}, we evaluate on CAMEO 2022 and the PDB Date Split~\cite{campbell2024generative}, and quantify prediction accuracy by comparing the generated structures against their ground-truth natural structures with RMSD and TM-score.

\textbf{Inverse folding} generates amino acid sequences compatible with a specified target structure~\citep{dauparas2022robust,hsu2022learning}. Following established protocols~\cite{wang2024dplm,hsieh2025elucidating}, we evaluate on CAMEO 2022 and the PDB Date Split. To assess structural self-consistency, we refold each generated sequence with ESMFold~\citep{lin2023esm2} and compute scTM relative to the target structure. As multiple sequences can yield about the same backbone structure, the amino acid recovery rate (AAR) is not a direct quality metric for inverse folding. Nevertheless, we report it to distinguish structural consistency from native-sequence recovery.

In our experiments, if no special indication is made, \textbf{every reported number is aggregated across five independent runs with distinct random seeds}, formulated as ``mean ± std'' or ``mean (min, max)''.
The RMSD and TM-score are calculated using standard functions in OpenFold~\citep{ahdritz2024openfold} and TM-Tools~\citep{zhang2005tm}.
The secondary structure is annotated via Biotite~\cite{kunzmann2023biotite}.
The number of clusters is obtained by clustering the generated structures via Foldseek~\citep{van2024fast}, using this command:
{
\setlength{\abovedisplayskip}{5pt}
\setlength{\belowdisplayskip}{5pt}
\setlength{\abovedisplayshortskip}{3pt}
\setlength{\belowdisplayshortskip}{3pt}
\setlength{\jot}{3pt}
\begin{align*}
    &\texttt{foldseek easy-cluster } \langle \texttt{input\_path} \rangle \\[-2pt] 
    &\langle \texttt{output\_path} \rangle \ \langle \texttt{tmp\_path} \rangle \\[-2pt]
    &\texttt{--alignment-type 1 --cov-mode 0} \\[-2pt]
    &\texttt{--min-seq-id 0 --tmscore-threshold 0.5}.
\end{align*}
}

\subsection{Details in Vanilla Sampling}
\label{asec:vanilla_sampling}

\subsubsection{Default Sampling Strategies}
\label{asec:default_sampling}

\begin{table*}[t!]\scriptsize
\centering
\captionsetup{font=small,skip=3pt}
\caption{Default Sampling Strategies for Unconditional Co-Generation}
{
\begin{tabular}{ccccc}
\toprule
& DPLM-2 & DPLM-2.1 & ESM3 & ESM3 \\
\midrule
Multimodal Order & Synchronous & Synchronous & Seq $\rightarrow$ Struct & SS $\rightarrow$ Struct $\rightarrow$ Seq \\
Sampling Step $T$ & 500 & 500 & $L$; $1$ & $L$ \\
Unmasking & Stochastic & Stochastic & Random & Random \\
Remasking & Enabled & Enabled & Disabled & Disabled \\
Temp. \& Annealing & From 2.0 to 0.1 & From 1.1 to 0.1 & From 1.0 to 0.0; 0.0 & 0.7 \\
\bottomrule
\end{tabular}
}
\label{tab:default_uncon_cogen}
\vspace{-6pt}
\end{table*}

\begin{table}[t!]\scriptsize
\centering
\captionsetup{font=small,skip=3pt}
\caption{Default Sampling for Motif-Scaffolding}
{
\setlength{\tabcolsep}{3pt}
\begin{tabular}{cccc}
\toprule
& DPLM-2 & DPLM-2.1 & ESM3 \\
\midrule
Multimodal Order & Synchronous & Synchronous & Seq $\rightarrow$ Struct \\
Sampling Step $T$ & 500 & 500 & $L/2$; $L/8$ \\
Unmasking & Stochastic & Stochastic & Random \\
Remasking & Enabled & Enabled & Disabled \\
Temp. \& Annealing & From 2.0 to 1.0 & From 1.1 to 0.1 & From 1.0 to 0.0 \\
\bottomrule
\end{tabular}
}
\label{tab:default_scaffolding}
\end{table}

\begin{table}[t!]\scriptsize
\centering
\captionsetup{font=small,skip=3pt}
\caption{Default Sampling for Structure Prediction}
{
\setlength{\tabcolsep}{3pt}
\begin{tabular}{cccc}
\toprule
& DPLM-2 & DPLM-2.1 & ESM3 \\
\midrule
Sampling Step $T$ & 100 & 100 & 1 \\
Unmasking & Deterministic & Deterministic & Deterministic \\
Remasking & Enabled & Enabled & Disabled \\
Temp. \& Annealing & 0.0 & 0.0 & 0.0 \\
\bottomrule
\end{tabular}
}
\label{tab:default_folding}
\end{table}

\begin{table}[t!]\scriptsize
\centering
\captionsetup{font=small,skip=3pt}
\caption{Default Sampling for Inverse Folding}
{
\setlength{\tabcolsep}{3pt}
\begin{tabular}{cccc}
\toprule
& DPLM-2 & DPLM-2.1 & ESM3 \\
\midrule
Sampling Steps $T$ & 100 & 100 & 8 \\
Unmasking & Deterministic & Deterministic & Random \\
Remasking & Enabled & Enabled & Disabled \\
Temp. \& Annealing & 0.0 & 0.0 & From 1.0 to 0.0 \\
\bottomrule
\end{tabular}
}
\label{tab:default_inv_folding}
\end{table}

We document the default sampling strategies for each base model across the foundational tasks considered in this work. For DPLM-2~\cite{wang2024dplm} and DPLM-2.1~\citep{hsieh2025elucidating}, we follow the official guideline and default hyperparameters\footnote{https://github.com/bytedance/dplm}. For ESM3~\citep{hayes2025simulating}, we adhere to the implementation details described in the original appendix, official codebase\footnote{https://github.com/evolutionaryscale/esm}, as well as established community practices~\cite{yim2025hierarchical,zhou2025hd}.
Tables~\ref{tab:default_uncon_cogen}-\ref{tab:default_inv_folding} present the default sampling configurations that serve as our primary baseline. These include the total sampling step, unmasking strategy, remasking allowance, sampling temperature, temperature annealing, and the multimodal sampling order. 

Examining the default configurations of DPLM-2/-2.1 and ESM3 across four foundational tasks reveals two implicit heuristics: 
(1) the required diffusion steps follow the order unconditional co-generation $\geq$ motif scaffolding $>$ structure prediction $\approx$ inverse folding; and 
(2) the randomness during sampling scales as unconditional co-generation $\approx$ motif scaffolding $>$ structure prediction $\approx$ inverse folding. 
We aim to examine these task-dependent assumptions and derive more nuanced insights. Furthermore, the inference protocols of DPLM-2/-2.1 and ESM3 exhibit significant implementation mismatches. 
The original ESM3 codebase lacks native support for stochastic unmasking, remasking, and synchronous sequence-structure co-sampling. Resolve this, we hope to standardize the ESM3 inference protocol and rigorously evaluate the effectiveness of each vanilla sampling strategy under a unified framework.

In addition to these, we retain several default configurations after careful consideration. First, to preserve each model's inherent training-inference alignment, we maintain the original diffusion schedules (linear for DPLM-2 and DPLM-2.1, cosine for ESM3) and temperature annealing functions (linear for DPLM-2 and DPLM-2.1, quadratic for ESM3). 
Second, for unmasking, DPLM-2 and DPLM-2.1 evaluate positional confidence using log-probabilities, whereas ESM3 relies on entropy, which has been confirmed in prior work to have little impact~\cite{lu2025towards}. 
Although these settings introduce minor methodological differences, they should not detract from our core investigation into the exploration-exploitation tradeoff.

\subsubsection{Optimal Vanilla Sampling Strategies}
\label{asec:optimal_vanilla_sampling}

Tables~\ref{tab:vanilla_uncon_cogen}-\ref{tab:vanilla_inv_folding} present the optimal vanilla sampling configurations we identified. Configurations that deviate from the defaults are highlighted in \diff{blue}. 

\begin{table}[t!]\scriptsize
\centering
\captionsetup{font=small,skip=3pt}
\caption{Optimal Vanilla Sampling for Uncond. Co-Gen.}
{
\setlength{\tabcolsep}{3pt}
\begin{tabular}{cccc}
\toprule
& DPLM-2 & DPLM-2.1 & ESM3 \\
\midrule
Multimodal Order & Synchronous & Synchronous & \diff{Synchronous} \\
Sampling Step $T$ & 500 & 500 & \diff{$L$} \\
Unmasking & Stochastic & Stochastic & \diff{Stochastic} \\
Remasking & Enabled & Enabled & \diff{Enabled} \\
Temp. \& Annealing & From 2.0 to 0.1 & From 1.1 to 0.1 & \diff{From 1.0 to 0.0} \\
\bottomrule
\end{tabular}
}
\label{tab:vanilla_uncon_cogen}
\vspace{-6pt}
\end{table}

\begin{table}[t!]\scriptsize
\centering
\captionsetup{font=small,skip=3pt}
\caption{Optimal Vanilla Sampling for Motif-Scaffolding}
{
\setlength{\tabcolsep}{3pt}
\begin{tabular}{cccc}
\toprule
& DPLM-2 & DPLM-2.1 & ESM3 \\
\midrule
Multimodal Order & Synchronous & Synchronous & \diff{Synchronous} \\
Sampling Step $T$ & \diff{$L$} & 500 & \diff{$L$} \\
Unmasking & Stochastic & Stochastic & Random \\
Remasking & Enabled & Enabled & \diff{Enabled} \\
Temp. \& Annealing & \diff{From 0.5 to 0.0} & \diff{From 0.5 to 0.0} & \diff{From 0.7 to 0.0} \\
\bottomrule
\end{tabular}
}
\label{tab:vanilla_scaffolding}
\vspace{-6pt}
\end{table}

\begin{table}[t!]\scriptsize
\centering
\captionsetup{font=small,skip=3pt}
\caption{Optimal Vanilla Sampling for Structure Prediction}
{
\setlength{\tabcolsep}{3pt}
\begin{tabular}{cccc}
\toprule
& DPLM-2 & DPLM-2.1 & ESM3 \\
\midrule
Sampling Step $T$ & \diff{$L/8$} & \diff{8} & \diff{8} \\
Unmasking & \diff{Random} & Deterministic & \diff{Stochastic} \\
Remasking & Enabled & Enabled & \diff{Enabled} \\
Temp. \& Annealing & \diff{From 0.1 to 0.0} & \diff{From 0.1 to 0.0} & \diff{From 0.5 to 0.0} \\
\bottomrule
\end{tabular}
}
\label{tab:vanilla_folding}
\end{table}

\begin{table}[t!]\scriptsize
\centering
\captionsetup{font=small,skip=3pt}
\caption{Optimal Vanilla Sampling for Inverse Folding}
{
\setlength{\tabcolsep}{3pt}
\begin{tabular}{cccc}
\toprule
& DPLM-2 & DPLM-2.1 & ESM3 \\
\midrule
Sampling Step $T$ & \diff{1} & \diff{8} & \diff{8} \\
Unmasking & \diff{Random} & \diff{Stochastic} & \diff{Deterministic} \\
Remasking & Enabled & Enabled & \diff{Enabled} \\
Temp. \& Annealing & \diff{From 0.1 to 0.0} & \diff{From 0.1 to 0.0} & \diff{From 0.5 to 0.0} \\
\bottomrule
\end{tabular}
}
\label{tab:vanilla_inv_folding}
\end{table}

\subsubsection{Ablation Study}
\label{asec:vanilla_sampling_investigation}

\begin{figure*}[b!]
    \centering
    \includegraphics[width=\linewidth]{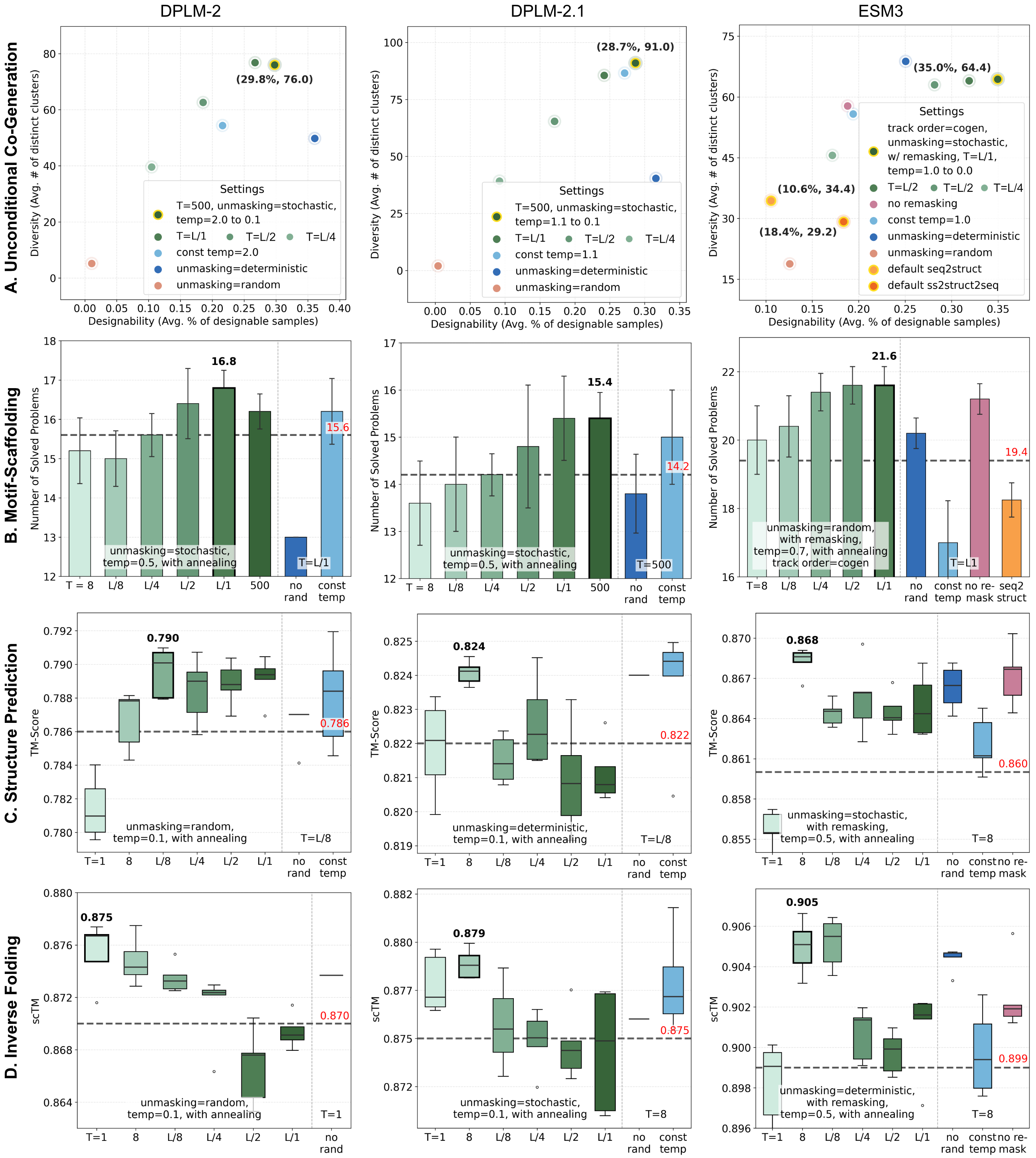}
    \captionsetup{font=small, skip=6pt}
    \caption{
    \textbf{Vanilla sampling: optimal configurations and ablation results}. 
    \textbf{A.} Pareto comparison between designability and diversity in unconditional protein sequence-structure co-generation. The first legend shows the optimal configuration, and each subsequent legend changes one setting at a time.
    \textbf{B-D.} The green bars to the left of the vertical dotted line show ablations over the total sampling step $T$, with the other vanilla sampling settings listed. 
    The remaining bars show one-at-a-time changes to the optimal configuration: 
    "no rand" denotes deterministic unmasking with temperature $= 0.0$ (argmax sampling); 
    "const temp" and "no remasking" denote disabled temperature annealing and remasking, respectively.
    }
    \label{afig:vanilla}
\end{figure*}

Due to space constraints, we present the comprehensive vanilla sampling ablation results across models and tasks in Figure~\ref{afig:vanilla}. 
Each panel reports task-specific metrics across varying sampling hyperparameters for a single base model, with the default-setting performance marked for reference.
The caption explains how to read the figure.

\subsection{Details in Reward-free Guidance}
\label{asec:cfg}

\subsubsection{Hyperparameter Selection}
\label{asec:cfg_hyperparameters}

\begin{table*}[t!]\scriptsize
\centering
\captionsetup{font=small,skip=3pt}
\caption{Selected CFG scales.}
{
\begin{tabular}{ccccc}
\toprule
& Uncon. Co-Gen. & Motif-Scaffolding & Structure Prediction & Inverse Folding \\
\midrule
DPLM-2 & \uline{$w_s=2.0$, $w_z=1.5$} & $w_s=1.5$, $w_z=1.0$ & $w_z=1.5$ & \uline{$w_s=4.0$} \\
DPLM-2.1 & \uline{$w_s=2.0$, $w_z=1.5$} & $w_s=1.5$, $w_z=1.0$ & $w_z=1.5$ & \uline{$w_s=2.0$} \\
ESM3 & \uline{$w_s=2.0$, $w_z=3.0$} & \uline{$w_s=2.0$, $w_z=2.0$} & \uline{$w_z=2.0$} & \uline{$w_s=2.0$} \\
\bottomrule
\end{tabular}
}
\label{atab:cfg_scale_employed}
\end{table*}

\begin{figure*}[t!]
    \centering
    \includegraphics[width=0.75\linewidth]{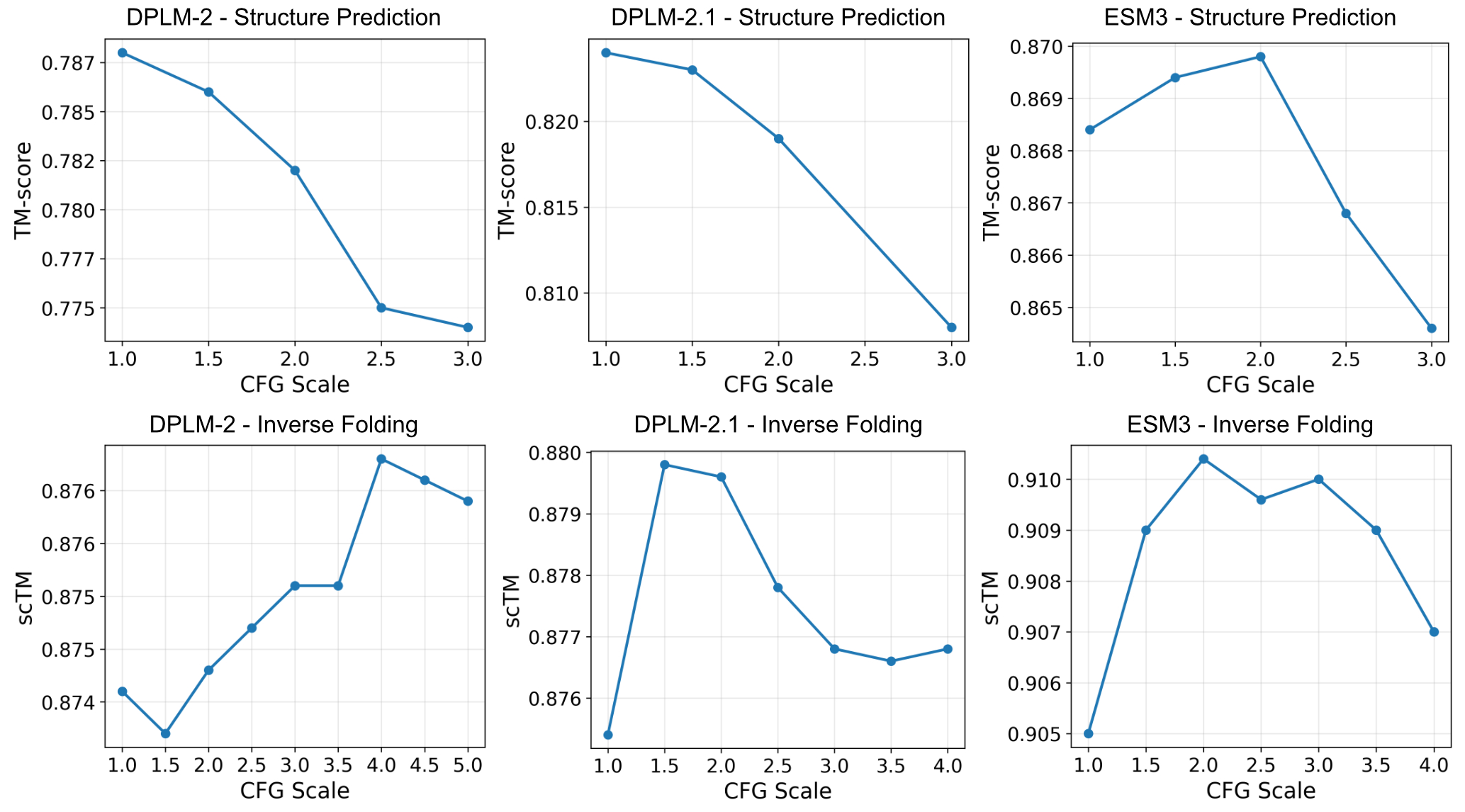}
    \captionsetup{font=small, skip=6pt}
    \caption{
    CFG scale analysis for protein structure prediction and inverse folding on CAMEO2022.
    }
    \label{afig:cfg_scales}
\end{figure*}
To select the CFG scales, we have conducted a limited hyperparameter search for each base model across the four tasks. Table~\ref{atab:cfg_scale_employed} lists the selected scales, with those yielding gains over the optimal vanilla sampling \uline{underlined}, and the corresponding results are reported in the main text.

A notable observation from this hyperparameter search concerns the sequence-structure asymmetry in multimodal pLMs. 
Since the two tracks are inherently imbalanced in tokenization, some differences in behavior are expected. 
However, the asymmetry in CFG response is substantially more pronounced for DPLM-2/-2.1 than for ESM3, as reflected in the contrast between protein structure prediction and inverse folding in Figure~\ref{afig:cfg_scales}. 
For DPLM-2 and DPLM-2.1, CFG provides no benefit for protein structure generation but yields clear gains when guiding inverse folding with a selected guidance scale. 
By contrast, ESM3 can positively respond to CFG in both directions, behaving more balanced across the two tasks.

A plausible explanation is that DPLM-2 and DPLM-2.1 are developed through modality extension, from sequence-only to multimodal pLMs, so the two tracks may not be internalized equally well. 
ESM3, trained natively at a larger scale on multimodal protein data, appears to internalize sequence and structure conditions more evenly, leading to a more balanced response to guidance across the two tracks. 
This is consistent with our insight that the exploitation effect of CFG depends on how well the base model has internalized the corresponding conditional signal.

\subsubsection{PDB Alignment Analysis on Unconditional Co-Generation Proteins}
\label{asec:pdb_tm_uncon_cogen}

To further confirm the effect of CFG in making ESM3's unconditional co-generation more "protein-like", we align the generated samples with PDB proteins. Specifically, we search each generated sample against the PDB database using Foldseek~\citep{van2024fast} and recorded the highest TM-score from the alignment against all PDB proteins, denoted as PDB-TM. 

In our benchmark of 500 generated proteins, ESM3 in the optimal vanilla sampling configuration yields an average of 387.2 samples that could be aligned to PDB proteins, with a mean PDB-TM of 0.548 ± 0.316. In contrast, ESM3 with cross-modal CFG yields an average of 473.0 samples with aligned PDB proteins, with a mean PDB-TM of 0.821 ± 0.156. These results demonstrate that cross-modal CFG makes ESM3's generations structurally closer to PDB proteins.

The concrete Foldseek command is: 
{
\setlength{\abovedisplayskip}{5pt}
\setlength{\belowdisplayskip}{5pt}
\setlength{\abovedisplayshortskip}{3pt}
\setlength{\belowdisplayshortskip}{3pt}
\setlength{\jot}{3pt}
\begin{align*}
    &\texttt{foldseek easy-search } \langle \texttt{input\_path} \rangle \\[-2pt] 
    &\langle \texttt{database\_path} \rangle \ \langle \texttt{output\_path} \rangle \\[-2pt]
    & \langle \texttt{tmp\_path} \rangle \ \texttt{--exhaustive-search} \\[-2pt]
    &\texttt{--alignment-type 1 --tmscore-threshold 0.0} \\[-2pt]
    &\texttt{--format-output query,target,qtmscore}.
\end{align*}
}

\subsubsection{Ablation Study on Motif-Scaffolding CFG}
\label{asec:scaffolding_cfg_ablation}

To validate our dual-condition CFG design for motif scaffolding, we compare it against two single-condition variants on ESM3. In all cases, the conditional logits remain those under full conditioning, namely $\mathbf{l}^{s,(t)}_{\text{cond}}, \mathbf{l}^{z,(t)}_{\text{cond}} = f_\theta(\bm{s}^{(t)}, \, \bm{z}^{(t)})$. For \uline{motif-only} CFG, the unconditional branch drops only the motif condition while retaining cross-modal context, yielding
\begin{align*}
    \mathbf{l}^{s,(t)}_{\text{uncond}}, \mathbf{l}^{z,(t)}_{\text{uncond}} = f_\theta\big(\bm{s}^{(t)}\setminus\bm{s}_{\text{motif}}, \, \bm{z}^{(t)}\setminus\bm{z}_{\text{motif}}\big).
\end{align*}
For \uline{cross-modal} CFG, the unconditional branch drops only the opposite-track context while retaining motif information, yielding
\begin{align*}
    \mathbf{l}^{s,(t)}_{\text{uncond}} = f_\theta\big(\bm{s}^{(t)}, \, \varnothing\big), \quad
    \mathbf{l}^{z,(t)}_{\text{uncond}} = f_\theta\big(\varnothing, \, \bm{z}^{(t)}\big).
\end{align*}
Table~\ref{atab:guided_scaffolding} reports the ablation results. Both single-conditioning CFG variants improve over the optimal vanilla sampling, while the full dual-conditioning CFG achieves the strongest gains. These results show that explicit motif guidance and implicit cross-modal guidance are both useful and are most effective when combined.

\begin{table}[h!]\scriptsize
\centering
\captionsetup{font=small,skip=3pt}
\caption{Ablation study on motif-scaffolding CFG.}
{
\setlength{\tabcolsep}{3pt}
\begin{tabular}{lccc}
\toprule
& \#Solved / 24 & Success Rate & \#Clusters \\
\midrule
ESM3 $_\text{Vanilla}$ & 21.6 (21, 22) & 30.9 ± 1.1 \% & 110.0 (100, 124) \\
ESM3 $_\text{CFG (Motif-only)}$ & 22.6 (22, 23) & 35.3 ± 0.6 \% & 146.8 (127, 158) \\
ESM3 $_\text{CFG (Cross-Modal)}$ & 22.4 (21, 23) & 36.6 ± 1.4 \% & 155.0 (144, 166) \\
ESM3 $_\text{CFG (Both)}$ & 22.8 (22, 23) & 37.5 ± 1.0 \% & 179.4 (165, 198) \\
\bottomrule
\end{tabular}
}
\label{atab:guided_scaffolding}
\vspace{-6pt}
\end{table}

\subsubsection{A Detailed Protein Structure Prediction Observation}
\label{asec:detailed_explain}

As noted in Table~\ref{tab:guided_folding}, ESM3 exhibits an interesting divergence between RMSD and TM-score with sampling strategies adjusted. 
Moving from default sampling ($T=1$, deterministic unmasking, no remasking, temperature $=0.0$) to the optimized vanilla sampling ($T=8$, stochastic unmasking, remasking, temperature annealing from 0.5 to 0.0) degrades RMSD while also improving TM-score on both datasets.
Applying CFG on top of the selected vanilla sampling strategy mitigates this RMSD--TM-score divergence, improving both metrics simultaneously.

\begin{figure*}[h!]
    \centering
    \includegraphics[width=0.65\linewidth]{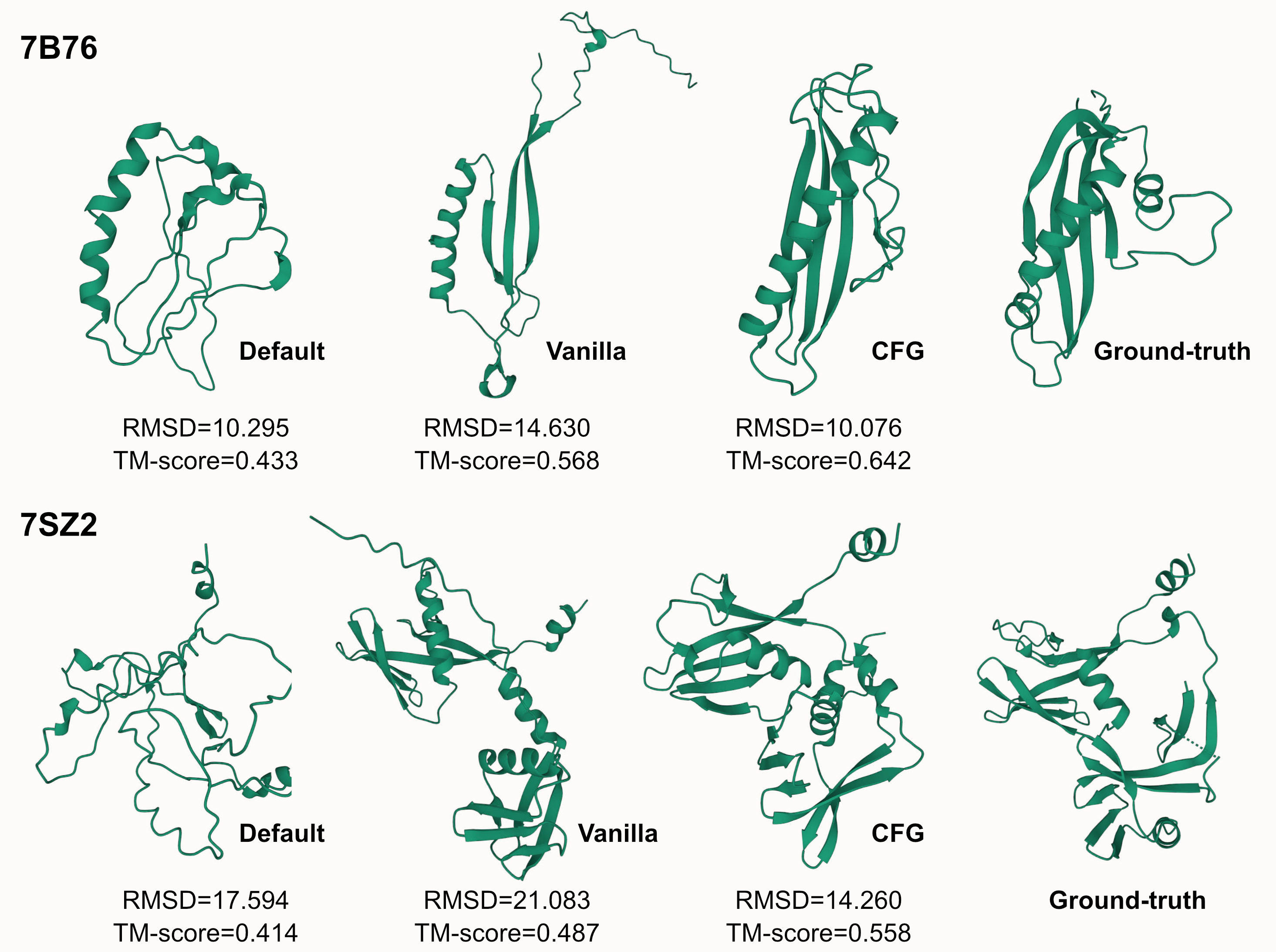}
    \captionsetup{font=small, skip=6pt}
    \caption{
    Two cases to illustrate ESM3's RMSD-TM-score divergence in protein structure prediction.
    }
    \label{afig:rmsd_tm_case}
\end{figure*}

Figure~\ref{afig:rmsd_tm_case} provides two case studies to illustrate this behavior. 
For some challenging targets, default single-pass argmax decoding yields low-quality predictions with incorrect secondary-structure organization. 
Multi-step iterative sampling recovers more plausible global topologies, which improves TM-score, but local fragment misorientations can persist and keep RMSD high. 
CFG further refines these local geometries, reducing this mismatch and yielding predictions with both more native-like global folds and more accurate local alignments.

\subsection{Details in Reward-guided Search}
\label{asec:rewarded_search}

\subsubsection{Implementation of Specialist Models}
\label{asec:rewarded_impl}

For unconditional protein sequence-structure co-generation, we run La-Proteina~\cite{geffner2025laproteina} using the pre-trained checkpoint\footnote{https://catalog.ngc.nvidia.com/orgs/nvidia/teams/clara/\\resources/ld2\_ucond\_tri\_512.ckpt/files} and follow the official guideline\footnote{https://github.com/NVIDIA-Digital-Bio/la-proteina}, using the default noise scales of 0.1 for the alpha-carbon atoms and 0.1 for the latent variables.
We also benchmark the official SVDD implementation in ESM3 following the official tutorial\footnote{https://github.com/evolutionaryscale/esm/blob/main/\\cookbook/tutorials/5\_guided\_generation.ipynb}. Given a desired length $L$, it employs a sequence-to-structure sampling order: sequence tokens are first sampled over $L/8$ steps with random unmasking, no remasking, and temperature annealing from 1.0 to 0.0, after which structure tokens are predicted in a single argmax step. On top of this default procedure, SVDD branches 10 candidates at each step and retains the one with the best pTM.
For motif scaffolding, we run La-Proteina using the motif-scaffolding checkpoint\footnote{https://catalog.ngc.nvidia.com/orgs/nvidia/teams/clara/\\resources/ld4\_motif\_idx\_aa.ckpt/files} with the default sampling settings.
For inverse folding, we run ProteinMPNN~\cite{dauparas2022robust} using the official checkpoint "v\_48\_020.pt" with temperature $= 0.1$.

\subsubsection{Beam Search for Multimodal pLMs}
\label{asec:beam_search_multimodal_plms}

\begin{algorithm}[t!]
\small
\caption{Generalized Beam Search}
\label{alg:beam}
\begin{algorithmic}[1]
\Require initial state $(\bm{s}^{(T)}, \bm{z}^{(T)})$, total steps $T$, beam width $N$, branching factor $B$, scoring interval $K$, reward $r(\cdot)$
\State $\mathcal{T} \gets \{(\bm{s}^{(T)}_n, \bm{z}^{(T)}_n)\}_{n=1}^{N}$ 
\For{$t = T, T{-}1, \dots, 1$}
    \If{$t \bmod K = 0$ \textbf{and} $t < T$}
        \State $\mathcal{C} \gets \textsc{Expand}(\mathcal{T}, B)$
        \State $\hat{r}(\bm{c}) \gets r(\textsc{QuickUnmask}(\bm{c})),\ \forall \bm{c} \in \mathcal{C}$
        \State $\mathcal{T} \gets \textsc{Select}(\mathcal{C}, \hat{r}, N)$
    \Else
        \State $\mathcal{T} \gets \textsc{Expand}(\mathcal{T}, 1)$
    \EndIf
\EndFor
\State \Return $\textsc{Select}(\mathcal{T}, r, 1)$
\end{algorithmic}
\end{algorithm}

In this study, we adapt beam search to the discrete diffusion sampling of multimodal pLMs, operating jointly over protein sequence and structure $(\bm{s}, \bm{z})$, as summarized in Algorithm~\ref{alg:beam}.
This procedure maintains $N$ trajectories in parallel in the set $\mathcal{T}$.

At each step $t$, the search layer calls $\textsc{Expand}(\cdot)$, which applies the underlying sampler for one step independently to each retained trajectory. 
At general steps, $\textsc{Expand}(\mathcal{T}, 1)$ simply advances each retained trajectory once.
Meanwhile, every $K$ steps is an expansion step to periodically expand, score, and prune trajectories: 
First, $\textsc{Expand}(\mathcal{T}, B)$ yields $N \times B$ candidate trajectories, collected in set $\mathcal{C}$.
Second, since partially masked candidates cannot be directly evaluated, we apply a single-forward argmax decoding $\textsc{QuickUnmask}(\cdot)$ to predict all remaining masked tokens and compute $r(\cdot)$ on the resulting proteins, leading to the estimated rewards $\hat r$.
Third, the selection rule $\textsc{Select}(\cdot)$ then chooses $N$ beams from $\mathcal{C}$ according to $\hat r$ for the next step of denoising.

After all $T$ steps, the final output is produced by applying the same $\textsc{Select}(\cdot)$ rule to the $N$ completed beams with target width $1$.
Notably, this formulation unifies two typical cases: $N{=}1, K{=}1$ recovers single-trajectory search with per-step reranking, i.e., SVDD~\cite{li2024derivative}, and $B{=}1, K{=}T$ reduces to Best-of-$N$ sampling.

Within this framework, the task-aware design choices in protein modeling reduce to two questions: how to instantiate the reward $r(\cdot)$, and how to instantiate the selection rule $\textsc{Select}(\cdot)$.
\begin{itemize}[leftmargin=*,itemsep=3pt,parsep=0pt]
\item \textbf{Reward}. While external models such as ESMFold could in principle serve as the scorer, we rely on signals internal to the multimodal pLM. 
This reflects our choice to keep all conditioning signals internal to the model, both to preserve the ``all-in-one'' nature of multimodal pLMs and to ensure a fair comparison with vanilla and guided sampling. 
As described in Section~\ref{asec:tokenization}, ESM3's protein structure tokenizer natively produces pTM scores during de-tokenization, which we use as global, trajectory-level rewards. A candidate on the structure track is scored directly (named structural pTM), whereas a candidate on the sequence track is first folded with an additional single-pass argmax decoding and then scored (named foldability pTM). 
\item \textbf{Selection}. The choice of $\textsc{Select}(\cdot)$ is aligned with the exploration-exploitation profile of each task. For the exploitation-biased and balanced tasks,
$\textsc{Select}(\cdot)$ simply returns the top-$N$ (or top-$1$) candidates ranked by the corresponding pTM: structural pTM for structure prediction, foldability pTM for inverse folding, and their sum for motif-scaffolding. For the exploration-biased task (unconditional co-generation), we optionally use a threshold-based random selection that uniformly samples from candidates satisfying both per-track criteria (foldability pTM $> 0.8$ and structural pTM $> 0.8$), which preserves diversity while enforcing a quality lower bound.
Notably, pTM $>0.8$ is a commonly used threshold~\cite{hayes2025simulating}.
\end{itemize}

\subsubsection{Hyperparameter Selection}
\label{asec:rewarded_hyperparameters}

In the reward-guided beam search, to select the beam width $N$, branching factor $B$, and scoring interval $K$, we conduct a limited hyperparameter search and obtain some primary observations.

\begin{table}[t]\scriptsize
\centering
\captionsetup{font=small,skip=3pt}
\caption{ESM3 $_\text{Beam}$ hyperparameter selection: unconditional co-generation}
{
\setlength{\tabcolsep}{3pt}
\begin{tabular}{cccccccc}
\toprule
& & \multicolumn{2}{c}{Designable Subset} & \multicolumn{4}{c}{All Samples} \\
\cmidrule(l){3-4} \cmidrule(l){5-8}
$N$ & $B$ & \#Design & \#Clusters & pLDDT & scTM & \#Clusters & $\alpha$ / $\beta$ \% \\ 
\midrule
2 & 1 & 318.0 & 137.0 & 89.529 & 0.955 & 194.0 & 45.7 / 12.8 \\
\rowcolor{gray!15}
\textbf{4} & \textbf{1} & 350.4 & 139.0 & 90.704 & 0.964 & 177.6 & 47.9 / 11.6 \\
8 & 1 & 342.0 & 132.0 & 90.632 & 0.963 & 175.0 & 47.6 / 11.7 \\
\midrule
4 & 2 & 385.0 & 119.0 & 91.342 & 0.973 & 138.0 & 50.2 / 10.5 \\
4 & 4 & 361.0 & 117.0 & 91.574 & 0.970 & 135.0 & 50.0 / 10.9 \\
4 & 8 & 382.0 & 107.0 & 91.251 & 0.973 & 135 & 50.7 / 10.4 \\
\bottomrule
\end{tabular}
}
\label{atab:rewarded_uncon_cogen_nb}
\end{table}

\begin{table}[t]\scriptsize
\centering
\centering
\captionsetup{font=small,skip=3pt}
\caption{ESM3 $_\text{Beam}$ hyperparameter selection: motif scaffolding}
{
\setlength{\tabcolsep}{3pt}
\begin{tabular}{lcccc}
\toprule
$N$ & $B$ & \#Solved & Success Rate & \#Clusters \\
\midrule
1 & 2 & 23.0 & 38.2 \% & 174.0 \\
1 & 4 & 23.0 & 38.0 \% & 186.0 \\
\midrule
2 & 1 & 23.0 & 43.7 \% & 222.0 \\
2 & 2 & 23.0 & 46.0 \% & 240.0 \\
2 & 4 & 23.0 & 47.8 \% & 233.0 \\
\midrule
\rowcolor{gray!15}
\textbf{4} & \textbf{1} & 23.0 & 49.9 \% & 261.0 \\
4 & 2 & 23.0 & 52.9 \% & 228.5 \\
4 & 4 & 23.0 & 54.0 \% & 221.0 \\
\bottomrule
\end{tabular}
}
\label{atab:rewarded_scaffolding_nb}
\end{table}

For unconditional cogeneration, we examine $N \in \{2, 4, 8\}$ and $B \in \{1, 2, 4, 8\}$ while fixing $K=T/5$. For motif scaffolding, we examine $N \in \{1, 2, 4\}$ and $B \in \{1, 2, 4\}$ while fixing $K=T/5$. The results are reported in Tables~\ref{atab:rewarded_uncon_cogen_nb} and \ref{atab:rewarded_scaffolding_nb}, respectively.
In both tables, only the selected (gray-highlighted) rows are aggregated over five random seeds, whereas other rows report single-seed results. 
Overall, increasing $N$ consistently improves designability, suggesting that a wider beam is more likely to preserve at least one high-quality trajectory for final selection. By contrast, increasing $B$ can reduce diversity, especially in a wide beam. Although a larger branching factor expands intermediate step exploration, it also leads to more aggressive pruning, so trajectories that might later yield diverse outcomes can be discarded too early, causing the search to concentrate on trajectories that appear better at the current step.

\begin{figure}[t]
    \centering
    \includegraphics[width=\linewidth]{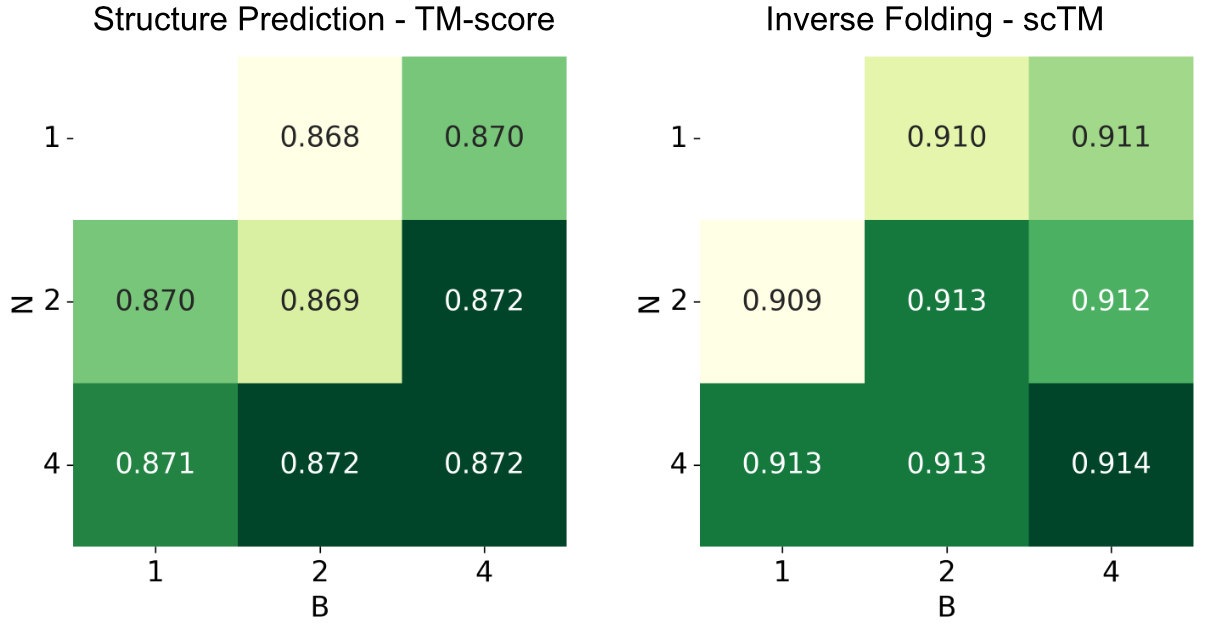}
    \captionsetup{font=small, skip=6pt}
    \caption{
    ESM3 $_\text{Beam}$ hyperparameter selection. Left: protein structure prediction; right: inverse folding.
    }
    \label{afig:reward_for_inv_folding_nb}
\end{figure}

For protein structure prediction and inverse folding, we examine $N \in \{1, 2, 4\}$ and $B \in \{1, 2, 4\}$ on the CAMEO 2022 dataset while fixing $K=1$, and report the results. As Figure~\ref{afig:reward_for_inv_folding_nb} shows, larger $N$ and $B$ consistently improve performance on both tasks. We attribute this trend to the fact that protein structure prediction and inverse folding are already relatively exploitation-biased. 
A larger beam width is more likely to retain high-quality trajectories, while a larger branching factor makes it easier to identify the best trajectory during search.

\subsubsection{Beam Search: Explore-then-Select}
\label{asec:explore_in_beam}

\begin{table}[t]\scriptsize
\centering
\captionsetup{font=small,skip=3pt}
\caption{ESM3 $_\text{Beam}$ explore-then-select: unconditional protein sequence-structure co-generation}
\label{atab:exploration_ablation_uncon_cogen}

\begin{subtable}{\columnwidth}
\centering
\captionsetup{font=scriptsize,skip=3pt}
\caption{Designable Subset}
{\setlength{\tabcolsep}{3pt}
\begin{tabular}{lcc}
\toprule
Temp. & \#Design $\uparrow$ & \#Clusters $\uparrow$ \\
\midrule
From 1.0 to 0.0 & 393.8 (383, 405) & 117.0 (107, 127) \\
1.0 & 350.4 (336, 360) & 139.0 (132, 145) \\
\bottomrule
\end{tabular}}
\end{subtable}

\vspace{3pt}

\begin{subtable}{\columnwidth}
\centering
\captionsetup{font=scriptsize,skip=3pt}
\caption{All Samples}
{\setlength{\tabcolsep}{3pt}
\begin{tabular}{lcccc}
\toprule
Temp. & pLDDT $\uparrow$ & scTM $\uparrow$ & \#Clusters $\uparrow$ & $\alpha$ / $\beta$ \% \\
\midrule
1.0 to 0.0 & 92.106 ± 4.781 & 0.974 ± 0.048 & 144.0 (137, 152) & 54.7 / 10.1 \\
1.0 & 90.704 ± 5.525 & 0.964 ± 0.056 & 177.6 (170, 186) & 47.9 / 11.6 \\
\bottomrule
\end{tabular}}
\end{subtable}
\end{table}

\begin{table}[t]\scriptsize
\centering
\centering
\captionsetup{font=small,skip=3pt}
\caption{ESM3 $_\text{Beam}$ explore-then-select: motif-scaffolding}
{
\setlength{\tabcolsep}{3pt}
\begin{tabular}{lccc}
\toprule
Temperature & \#Solved / 24 & Success Rate & \#Clusters \\
\midrule
From 0.7 to 0.0 & 23.0 (23, 23) & 48.9 ± 0.9 \% & 214.6 (202, 226) \\
From 1.0 to 0.0 & 23.0 (23, 23) & 49.9 ± 0.7 \% & 261.0 (250, 278) \\
\bottomrule
\end{tabular}
}
\label{atab:exploration_ablation_scaffolding}
\end{table}

\begin{table}[t!]\scriptsize
\centering
\captionsetup{font=small,skip=3pt}
\captionof{table}{ESM3 $_\text{Beam}$ explore-then-select: inverse folding}
{
\setlength{\tabcolsep}{3pt}
\begin{tabular}{lcccc}
\toprule
~ & \multicolumn{2}{c}{CAMEO} & \multicolumn{2}{c}{PDB Date Split} \\
\cmidrule(l){2-3} \cmidrule(l){4-5}
Umasking & scTM $ \uparrow $ & AAR & scTM $ \uparrow $ & AAR \\
\midrule
Deterministic & 0.910 ± 0.134 & 45.7 & 0.954 ± 0.075 & 49.1 \\
Stochastic & 0.914 ± 0.129 & 46.0 & 0.956 ± 0.069 & 49.3 \\
\bottomrule
\end{tabular}
}
\label{atab:exploration_ablation_inv_folding}
\end{table}

Across three of the four multimodal protein modeling tasks, beam search works better when the base sampler is made slightly more exploratory. 
For each task, we start from the selected vanilla sampling configuration (Table~\ref{tab:vanilla_uncon_cogen}-\ref{tab:vanilla_inv_folding}) and CFG scale (Table~\ref{atab:cfg_scale_employed}), and modify only one vanilla sampling setting.
Table~\ref{atab:exploration_ablation_uncon_cogen}-\ref{atab:exploration_ablation_inv_folding} report the ablation results.

For unconditional co-generation, replacing temperature annealing from 1.0 to 0.0 with a constant temperature of 1.0 trades designability for diversity: the number of designable samples drops from 393.8 to 350.4, while the number of clusters on the designable subset rises from 117.0 to 139.0. 
For motif scaffolding, increasing the initial temperature from 0.7 to 1.0 improves both success rate (48.9\% to 49.9\%) and diversity (214.6 to 261.0 clusters). 
For inverse folding, switching the unmasking strategy from deterministic to stochastic yields modest but consistent gains in scTM on both CAMEO 2022 and the PDB Date Split.

We suspect that, without this additional exploration, the beams remain too similar for reward-based selection to meaningfully distinguish among the protein generation candidates. 
When exploring a too-narrow candidate space, the search could fail to achieve substantial diversity gains in more open-ended, exploratory-biased tasks or self-consistency gains in exploitation-biased tasks.

\subsection{Inference Efficiency Analysis}
\label{asec:effciency_analysis}

As shown in Tables~\ref{tab:guided_uncon_cogen}-\ref{tab:guided_inv_folding}, in addition to benchmark performance, we report the specific FLOPs and actual wall-clock runtime on a unified device (a single Nvidia H20 GPU). 
Based on these results, we can draw the following analysis.

\noindent
\textbf{FLOPs and Forward Passes}. For benchmarks with a fixed number of tokens, FLOPs are roughly proportional to the number of forward passes. In the vanilla sampling stage, the number of passes equals the diffusion steps $T$. In the CFG stage, the model combines conditional and unconditional outputs, requiring 2 passes per step for structure prediction and inverse folding, and 3 passes for motif scaffolding and unconditional co-generation. That is, $2T$ and $3T$ passes in total, respectively. 

In the reward-guided search stage, let $N$ be the beam width, $B$ the branching factor, and $K$ the scoring interval. The number of branching (scoring) steps is $S=\lfloor(T-1)/K\rfloor$. The total forward passes equal $((T-S)N+SNB)$ $\times$ per-step cost (2 or 3 passes depending on the CFG formulation), plus an additional $(SNB+N)$ $\times$ reward cost (1 to 3 passes depending on the reward type). Specifically, our beam search inference requires $2((T-S)N+SNB)+(SNB+N)$ passes for structure prediction, $2((T-S)N+SNB)+2(SNB+N)$ passes for inverse folding, and $3((T-S)N+SNB)+3(SNB+N)$ passes for motif scaffolding and unconditional sequence-structure co-generation.

\noindent
\textbf{Runtime vs. FLOPs Scaling}. Due to GPU parallelization optimizations, the actual inference runtime often grows sublinearly with FLOPs. Depending on the base model implementation, if the GPU utilization is close to 100 percent, the runtime scales nearly proportionally with FLOPs. Conversely, if there is spare GPU capacity, the runtime may grow slowly even when FLOPs increase significantly.

\noindent
\textbf{Suboptimal Efficiency in Default Inference}. 
Default inference protocols are often suboptimal in both performance and efficiency.
Through our first-stage vanilla sampling exploration, we can achieve simultaneous runtime reduction and performance improvement for DPLM-2/2.1 (structure prediction, inverse folding) and ESM3 (unconditional co-generation, inverse folding). 
This is achieved by reducing the total number of pLM forward passes, i.e., setting fewer diffusion steps and enabling synchronous sequence-structure co-sampling.

\noindent
\textbf{Performance-Efficiency Balance in CFG Inference}. Our classifier-free guidance (CFG) strategies offer a superior performance-efficiency balance, delivering various degrees of performance improvement with only 1 to 3 times runtime overhead. 
Especially for ESM3, whose implementation leaves spare GPU capacity, resulting in only a moderate increase in runtime despite doubling the FLOPs, yielding practical benefits across all four tasks.
In the unconditional co-generation benchmark, ESM3 runtime increases by about 1.46 times, while the number of designable clusters increases by about 1.81 times. In the motif scaffolding benchmark, ESM3 runtime remains nearly flat, while the number of clusters increases by a factor of 1.63. 
Similarly, for structure prediction and inverse folding, ESM3 performance improves while runtime remains nearly constant ($<$1.1 times).

\noindent
\textbf{Beam Search Efficiency}. 
Table~\ref{tab:guided_uncon_cogen}-\ref{tab:guided_inv_folding} results acknowledge that reward-guided beam search significantly increases inference costs. Meanwhile, as shown in the FLOPs estimation formats, increasing the beam width, branching factor, or scoring frequency multiplies the computational overhead.
However, we believe this trade-off is justifiable for two reasons. First, our approach is training-free. Unlike specialized models or post-training studies that require multi-node clusters and weeks of training, we adjust sampling strategies using a single GPU for at most several hours. Second, the performance gains at this stage demonstrate that pLMs can compete with specialized state-of-the-art models, overturning prior consensus on multimodal pLMs such as ESM3.

\end{document}